\documentclass[%
 reprint,
 amsmath,amssymb,
 aps,
]{revtex4-2}

\usepackage{graphicx}% Include figure files
\usepackage{dcolumn}% Align table columns on decimal point
\usepackage{bm}% bold math
\usepackage{bbm}
\usepackage{array}

\usepackage{parskip}

\usepackage{tabularx}

\begin{document}

\preprint{APS/123-QED}

\title{Network analysis of students' drawn representations of an introductory lab}% Force line breaks with \\
% \thanks{A footnote to the article title}%

\author{W. Brian Lane$^{1,2}$}
%  \altaffiliation{Department of Physics, University of North Florida.}%Lines break automatically or can be forced with \\
 \email{Brian.Lane@unf.edu}
\author{Charlotte Dries$^{3}$}
\author{Gabriella Khazal$^{4}$}
% \author{Thomas O'Brien$^{1}$}
\author{Tiffany Snow$^{1}$}
\affiliation{%
 $^{1}$Department of Physics, University of North Florida\\
 $^{2}$Northeast Florida Center for STEM Education, University of North Florida\\
 $^{3}$Department of Biology, University of North Florida\\
 $^{4}$Department of Psychology, University of North Florida\\
 1 UNF Drive, Jacksonville, FL, 32224% \textbackslash\textbackslash
}%

\date{\today}% It is always \today, today,
             %  but any date may be explicitly specified

\begin{abstract}

Introductory physics labs can be designed and studied using the Communities of Practice framework, in which a group of members pursues a common set of goals by learning and implementing a set of agreed-upon practices. Studies of student preferences and behaviors in lab show how students might perceive the introductory lab community differently, as they occupy different positions within the community and engage differently with its practices. Such differences can be particularly sharp along the dimensions of gender, college generation, and racial background. In this proof-of-concept study, we demonstrate how we can explore students' perceptions of the introductory lab community of practice using a drawing-based survey and network analysis. The drawing-based survey collects students' open expressions of their perspectives and guides them to elaborate on the goals, members, and practices of the community. We catalog the distinct elements from these drawings and categorize them as related to goals, members, and practices. With this quantitative count of drawing elements, network analysis then allows us to obtain an overview of how the elements in these drawings relate to each other while preserving the openness of student expression. We are particularly interested in the frequency of each element's depiction across the network and its betweenness centrality within the network. We collected $N = 74$ student drawings from a studio-format introductory physics for life sciences sequence. Our network analysis reveals a wide diversity of drawing elements with a focus on centralized hands-on practices and community members and a sparsity of goals across the students' perspectives. We demonstrate how some elements show differences with large effect sizes between identity groups based on gender, college generation, and racial background. We compare these differences to those found in observational studies and discuss future uses of the drawing survey and network analysis approach.

\end{abstract}

%\keywords{Suggested keywords}%Use showkeys class option if keyword
                              %display desired
\maketitle

%\tableofcontents

\section{Introductory Labs as Communities of Practice}

Introductory labs are an important component of undergraduate STEM education that impact student persistence and STEM identity \cite{esparza2020characterization,kim2018developing,seyranian2018longitudinal}. These labs develop students' conceptual learning and experimental skills, and they enculturate students into STEM as a community-oriented profession based on cooperative work using shared resources \cite{holmes2016examining,irving2014conditions,dew2024group}. This enculturation is not merely a byproduct of the introductory lab experience, but an important learning goal as instructors model the broader STEM community in a local microcosm \cite{irving2014conditions}. For example, a collaborative lab experience can show students, who might express a preference to work alone, how their future employers or graduate programs expect them to work in a collaborative setting \cite{hood2021like,mckay2024student,clarke2018rethinking}. The structure of group work in an introductory lab can establish and reinforce norms for interacting with colleagues and communicating ideas \cite{hagvall2024students,ping2025group,wieselmann2020just,reinsvold2012power,krystyniak2007analysis}. An introductory lab modeled after authentic scientific inquiry can help expand students' conceptions of science beyond ``the scientific method'' \cite{woodcock2014scientific,tang2010scientific}. Physics education in particular has taken up the charge to enculturate students into using computational and experimental practices to create models, design experiments, analyze and visualize data, and communicate results and ideas \cite{kozminski2014aapt,AAPTRECS}. Such enculturation gives students context and motivation to assimilate and practice the concepts and experimental skills they develop in the introductory lab and prepares students for future participation in the broader STEM community in graduate programs, research labs, or industry.

To facilitate this enculturation in course design and research its effects on student learning, it is helpful to think of an introductory lab as a Community of Practice (COP), defined as a group of members pursuing a common set of goals by using conventional practices \cite{nicolini2022understanding,wenger1999communities,Wenger2000Communities,wenger2002cultivating,prefontaine2021informal,rosen2022working}. In the context of an introductory physics lab, members would include the instructor, a student's lab partners, and students in other lab groups; goals might include meeting course outcomes, getting a good grade, and exploring topics of personal relevance; and practices might include the use of lab equipment, data analysis techniques, modes of collaboration, and particular software. Learning in a COP is pictured as navigating a trajectory from peripheral membership toward central membership as a student aligns with these goals, adopts these practices, and begins to identify as a member of the COP \cite{irving2014conditions,quan2018interactions,wenger1999communities,sissi2011learning,irving2020communities,lave1991situating,li2013physics,close2016becoming,boylan2016deepening,boaler2000mathematics,nardi2003mathematics}. For a student in an introductory physics lab, this navigation might involve developing their competency with equipment use and data analysis alongside an increased sense of involvement arising from agency in decision-making \cite{wan2024characterizing}. Members of a COP often need to negotiate their membership across a nexus of multiple communities \cite{wenger1999communities,close2016becoming,irving2020communities,diekman2011malleability,diekman2010seeking,moshfeghyeganeh2021effect,khong2021community}. In the context of an introductory physics lab, this nexus might include the physics lab, introductory labs in their major, a study group, student clubs, and affinity groups. 

Irving, McPadden, and Caballero argue that a COP-informed course design offers ``space for identity development, align[s] with previous transform efforts, creates [a] learning environment of trust, respect, and risk taking, [and] facilitates a focus on practice development'' \cite{irving2020communities}. Such design attends to the trajectories students follow by engaging in legitimate peripheral participation, defined as practices that are legitimized by central members of the community and appropriately formulated for a newcomer. This approach also invests broader meaning in the concepts and skills being learned and the collaborative nature of many engaged learning practices. COP-informed course design focuses on creating an environment for learning to take place, and how that learning is relevant to the student's other community memberships, such as in their major program or future professional communities.

Observational studies of introductory labs reveal that students of differing gender, college generation, and racial background participate in and experience this enculturation differently depending on the lab format and context \cite{doucette2022students,dew2024group,holmes2022evaluating,quinn2020group,dew2022so,doucette2020hermione,day2016gender,wilson2013science,kalender2019gendered,hazari2013science}. Doucette, Clark, and Singh attribute some of these differences to the ``masculinized culture of physics'' \cite{doucette2022students}. Multiple studies have observed female students preferring and taking on a secretary-type role as note-taker or manager \cite{dew2024group,doucette2020hermione,day2016gender} while male students prefer and take on a role using the lab equipment or analyzing data \cite{dew2024group,holmes2022evaluating,quinn2020group,dew2022so,day2016gender}, with partner agreements producing more equitable equipment usage \cite{dew2024group}. Sundstrom \textit{et al}. \cite{sundstrom2022examining} found that lab groups interacted with other groups with differing frequency depending on their gender composition (all-male, all-female, mixed-gender). From a COP perspective, these differences mean that different student groups are experiencing the learning community---and therefore scientific encultration---differently.

We argue that student experiences of an introductory lab as a COP are mediated by their \textit{perceptions} of that COP. For example, when students of different identity groups express different preferences for roles in the lab \cite{dew2024group,holmes2022evaluating,doucette2020hermione}, they form these preferences based on what they perceive STEM to be like, how they expect to fit in, and what tasks they expect to be competent in. Then, as these students take on different roles within the lab \cite{doucette2022students,dew2024group,quinn2020group,dew2022so,doucette2020hermione,day2016gender}, their perceptions become reified, modified, or unmet \cite{shi2025impact,wilson2013science,kalender2019gendered,hazari2013science}. These perceptions then shape the students' decisions regarding further participation in STEM learning and a STEM career \cite{zwolak2017students,dou2016beyond,alaee2023analyzing,diekman2011malleability,franz2016experiences,agosto2008changing,diekman2010seeking}. 

In this paper, we consider student perceptions of the COP created in a studio-format introductory physics for life sciences (IPLS) sequence. Studio is an integrated guided inquiry learning environment in which lecture is deemphasized as a means of information transfer in favor of ``cooperative group problem solving, experiments, or answering questions that students may have'' \cite{commeford2021characterizing}. This cooperative learning involves student interactions within the lab group and between lab groups, facilitated by documenting their work on large whiteboards. Such cooperative learning enhances self-efficacy \cite{dou2016beyond} and persistence \cite{zwolak2017students,zwolak2018educational,brewe2009investigating}, which are features of navigating toward central membership within a COP. IPLS is a movement of curricular reforms that recontextualizes the traditional algebra-based physics course in topics and problems relevant to life science majors (biology, kinesiology, pre-medical) \cite{crouch2023introductory,chessey2023living}. In an IPLS course, emphasis is placed on physics content and experimental skills most relevant to these students' salient interests in the life sciences \cite{crouch2014introductory,lapidus2020physics}. IPLS reforms have been shown to positively influence student interest and performance in physics \cite{crouch2018life}. We consider the IPLS studio sequence an important COP to study: The students are already members of a community together in their major programs, and IPLS is designed to emphasize the relevance of physics content and experimental skills they learn in the introductory physics community for their major's community. An IPLS course builds on their knowledge, practices, and membership formed within their major community, emphasizing authentic scientific practices and skills across disciplines. Finally, the studio format explicitly guides these students in legitimate peripheral participation so that they adopt practices and goals from the physics context.

To study students' perceptions of this COP, we administered an open-ended survey in which students draw and annotate depictions of their perceptions of IPLS studio. This drawing-based survey was validated using the response process evaluation method \cite{wolf2022survey} as recently described in \cite{lane2024using}. We studied themes and differences in these drawings using network analysis, which is rising in popularity across physics education research (PER), particularly in applications to qualitative data sets \cite{traxler2024person}. We share this work as a proof of concept that addresses the following research questions about these emerging methodologies in PER:

\textbf{RQ1}: How can network analysis techniques be used to represent and analyze qualitative data from students' drawn expressions of their perceptions of an introductory lab?

\textbf{RQ2}: What can network analysis show about students' perceptions of a studio-format IPLS course as a community of practice?

In answering these questions, we use the COP framework as an ideal to compare against the students' experiences. While instructors might wish to establish a shared sense of membership, practices, and goals, examining the student perspective could help us explore the degree to which that ideal has been achieved, and what future work could be done.

This study builds on our prior work describing students' mental models of a COP, which we review in Section \ref{sec:COP-models}. In Section \ref{sec:methods}, we review the two components of our methodology: a drawing-based survey that collects information about students' COP models, and network analysis techniques that we apply to the survey responses. In Section \ref{sec:authors}, we discuss how our identities and contexts as researchers have informed and shaped this work. In Section \ref{sec:results}, we review prominent features of the entire network of student drawings and differences exhibited in the drawings made by identity groups based on gender, college generation, and race. In Section \ref{sec:discussion}, we discuss how these differences compare with observations in previous studies. In Section \ref{sec:limitations}, we discuss the limitations of this study. In Section \ref{sec:implications}, we discuss implications for future research and instruction. Finally, in Section \ref{sec:conclusions}, we present our conclusions.

\section{COP Models: A Framework for Student Perceptions\label{sec:COP-models}}

A student's internal representation of their learning community guides decision-making and engagement in the learning process and participation in future education and career opportunities \cite{alaee2023analyzing,diekman2011malleability,franz2016experiences,agosto2008changing,diekman2010seeking}. Our group's prior work has demonstrated that these internal representations can have observable differences between students \cite{lane2022student} and between students and instructors \cite{lane2021comparison}. We attend to these internal representations through the construct of a mental model of a Community of Practice, or COP model. A student’s COP model includes the members of the COP, the institutional structures relevant to the COP, the student’s understanding of the COP's goals and practices, and the student's sense of membership (central or peripheral positioning) within the COP.

In particular, our previous work \cite{lane2022student} outlined how...

\begin{itemize}

\item A learner's COP model includes the goals and practices that make up their understanding of the COP's sense of joint enterprise \cite{irving2014conditions,close2016becoming,irving2020communities}. As the learner's interactions with the COP shape their identity \cite{sissi2011learning}, the COP model answers the question, ``What does the learner believe that they are interacting with?''

\item A learner's COP model includes their sense of membership and trajectory within the COP \cite{fracchiolla2020community,quan2018interactions}, guiding them like a map \cite{kelly2016social}. 

\item A learner develops their COP model in response to legitimate peripheral participation and feedback, forming their sense of the importance of those practices \cite{franz2016experiences,boylan2016deepening,irving2020communities,agosto2008changing,diekman2010seeking,diekman2011malleability}. 

\item A learner's COP model enables them to extrapolate their experience of a local COP to an understanding of a global COP \cite{quan2018interactions,knezek2013impact,christensen2013contrasts,shin2015changes,means2021cultivating}.

\item A learner can compare their models of different COPs they are a member of to negotiate their nexus of multimembership \cite{wenger1999communities,close2016becoming,irving2020communities,diekman2011malleability,diekman2010seeking,moshfeghyeganeh2021effect}.

\end{itemize}

COP models can provide a rich landscape to explore, including objects (the members, institutions, and items involved in the COP), descriptors (qualities of those objects relevant to the COP), behaviors (actions and practices used in the COP), interactions (including the goals that guide the members), and accuracy (the alignment between the learner's COP model and the COP in reality) \cite{louca2011objects}. Two students can experience the same community and develop different COP models, and these models can differ from the system in reality. A model might omit, oversimplify, or exaggerate features that exist in reality, or it might introduce new features. For example, our previous study of upper division physics majors learning computational physics for the first time found that these students prioritized different reasons for using computation \cite{lane2022student}. Similarly, our study of students engaged in undergraduate research found that biology majors expressed a greater sense of purpose for their research than physics majors did \cite{lane2024using}. 

To capture these differences, we developed a survey in which students externalize their COP model of an introductory lab by drawing \cite{lane2024using}. As described further in Section \ref{subsubsec:ilds}, the survey begins with an open-ended prompt for the student to draw a picture of what the introductory lab is like. The survey then presents a checklist of prompts for the student to depict the goals, members, and practices of the introductory lab's COP and how the student fits into this COP. This checklist helps students think along the dimensions of a COP without needing a lengthy introduction to the framework. Finally, the survey asks the student to write a verbal description of their drawing to support the researchers' identification of its contents. We discuss the theoretical foundation for this survey, its implementation, and our use of network analysis to study these drawings in the next section.

\section{Methodology: Drawing-Based Survey and Network Analysis\label{sec:methods}}

In this section we outline and justify the two methodological components of our study: drawing-based data collection and network analysis of drawing elements. A common motivation for choosing these methodologies is their ability to foreground students' expression of their perspectives.

\subsection{\label{sec:drawing}A Drawing-Based Survey for Introductory Labs}

We previously discussed the process of developing and validating our drawing-based survey in \cite{lane2024using}. Here, we summarize the foundations of this survey from the literature, present the introductory lab version of the survey used in this study, and share examples of how we thematically cataloged the drawings students created. 

\subsubsection{Why Drawings?\label{sec:WhyDrawing}}

While new to PER, drawing-based data collection has established uses in education research. We find the following features of drawing-based data collection particularly salient to our study of student perceptions of IPLS studio.

First, drawings can reveal differences in the perspectives that different populations hold \cite{uccar2023picturing,haney2004drawing}. Examples include studies about student perspectives of the learning experience \cite{berti2022draw,rott2023mathematics,hatisaru2022knowledge,brown2013illustrating,mckillop2007drawing} and student conceptions of what scientists and science are like \cite{symington1990draw,finson2002drawing,miller2018development,mccarthy2015teacher,meyer2019draw}. Because drawing is a ubiquitous activity that can bridge communication differences, drawing-based data can be used to compare students' perspectives across backgrounds and cultures \cite{picker2000investigating,oistein2023stereotypical,chambers1983stereotypic,hatisaru2022knowledge,cox1999children,chambers1983stereotypic}. Such insight is important in our study of students' internalized representations of an introductory lab as a community.

Second, the richness of drawing gives students a broad means of expressing their perspectives \cite{peterson2021speaking,sondergaard2019drawing,frederiksen2015mixed} while requiring minimal prompting from the researcher \cite{literat2013pencil,leavy2018introduction,mannay2010making,driessnack2012arts,cruwys2016social}. Such a balance is important in our study of students' perspectives since we want students to freely express their ideas without overly suggesting a specific type of response. We have struck this balance by starting our drawing-based survey with an open-ended starting prompt, followed by a checklist of items for students to consider that align with our framework, and then a prompt for a written explanation of the drawing's contents to guide our interpretation and analysis \cite{ehrlen2009drawings,bentley2020social,kuttner2017draw,mehlmann2017looking,fan2023drawing,frederiksen2015mixed,driessnack2012arts}. Therefore, we can collect data that aligns with an existing research framework \cite{lamminpaa2023draw,cvencek2014developing,sondergaard2019drawing,merriman2006using,lewis2003sage}, even when that framework invokes abstract elements, as drawing helps the student to express abstract concepts through symbolic forms \cite{martikainen2023drawing,thordsen1991comparison}. 

Third, this foregrounding of open expression engages students' perception, memory, and social cognition \cite{fan2023drawing} in ways that are relevant to them and their cultural background. This empowerment of expression encourages equitable involvement and promotes students' rights throughout the research process \cite{gameiro2018drawingout,literat2013pencil,merriman2006using}. 

Finally, this approach offers the pragmatic benefit that STEM majors are already trained to represent their thinking graphically \cite{lowe1993constructing,bongers2020building,finson2011visual,ehrlen2009drawings}%, including recent innovations in graphical abstracts \cite{lee2023current}
. We find that asking STEM students to draw about their experiences requires little prompting or follow-up, and most are eager to provide a wealth of detail about their perspectives, even when using simpler drawing styles such as stick figures \cite{reinisch2017methodical,losh2008some}. % For example, in the sample drawings we present in Figure \ref{fig:sample-drawings}, we see students using examples of representations learned in their science labs (graphs, velocity vectors, color-coded VPython text).

One important caution to keep in mind with drawing-based data collection is its potentially high load on visual and physical ability. As described by \cite{scanlon2019ability}, it is important for researchers to consider which dimensions of ability a research tool places high load on, potentially underserving students who experience limitations along those dimensions. One advantage to qualitative research is the room it affords for diverse and flexible methodologies, and McPadden, \textit{et al}. advocate for using planning tools to make such flexibility an explicit part of the research design \cite{mcpadden2023planning}. As we discuss in Section \ref{sec:implications}, we think the network analysis process presented in this section can be used to analyze a diversity of qualitative data sources, potentially allowing comparison between the data found in, say, drawings, interviews, and written surveys that each create demand on different dimensions of ability.

\subsubsection{Introductory Lab Drawing Survey\label{subsubsec:ilds}}

Our introductory lab drawing survey is a modified version of a research group drawing survey we previously shared in \cite{lane2024using}. The primary modification is the context (introductory physics studio instead of a research group) presented in the survey items, which are otherwise worded the same. As with the research group survey \cite{lane2024using}, we used the response process evaluation method \cite{wolf2022survey} to validate the introductory lab survey and confirmed it captures information about students' COP models with no further revision required. The introductory lab drawing survey reads as follows:

\begin{enumerate}

\item On a sheet of paper or a digital whiteboard, draw a picture that shows your experiences learning in the studio for Introductory Physics for Life Sciences (this course). Feel free to draw multiple scenes to communicate your thoughts, and show what it’s like working directly with others and working individually in this studio. In your picture, use labels or descriptions to identify who is there, what they are doing, and why.

\item Look over your picture and add the following if they’re missing from your picture:

$\square$ The goals people in your lab are trying to accomplish.

$\square$ Activities the people in your lab do to reach those goals.

$\square$ How these activities help accomplish these goals.

$\square$ How you, individually, participate in these activities and goals.

$\square$ How you feel about these activities and goals

\item In the space below, describe what’s in your picture, and give a few examples of events that inspired your picture.

\item Who do you typically work with (or who did you typically work with) in studio?

\item \label{genderq} What gender do you identify as? (Provide as much info as you like.)

\item How would you describe your race/ethnicity? (Provide as much info as you like.)

\item \label{majorq} What is your major?

\item Are you a first-generation college student?

\end{enumerate}

We formatted demographic questions \ref{genderq} through \ref{majorq} as free-response so students could report as much detail as they preferred. We then grouped demographic responses together when making subnetworks (for example, one student identified as Filipino and was included with all who identified as Asian).

We collected survey responses from the studio-format IPLS sequence (both Physics I and Physics II) near the end of the spring 2024 semester so that students had plenty of experience to form a mental model of the studio environment. We spent the following semester collaboratively cataloging these drawings using a checklist of questions \cite{lane2024using} about the objects, descriptors, behaviors, and interactions depicted \cite{louca2011objects}. This checklist required us to review each drawing as a whole four times and ensure we had named each element in the drawing. We then entered these elements into a code book to catalog similar elements across the full set of drawings. Each row in the code book lists a drawing element, and each column records whether each element was identified in a particular drawing. In this code book, we categorized each element as related to a goal, member, or practice of the lab's COP. We found that collaboration across a diversity of team members was key to the cataloging process as faculty (WBL) and students (CD, GK, TS) offer complementary insights into the experience of an introductory physics lab. As we entered more drawings' elements into the code book, we found we needed to add progressively fewer new elements, indicating a consistency across the drawings' contents. We also negotiated to find consensus on when to combine elements. For example, we quickly found it unnecessary to track individual items of lab equipment (carts, coffee filters, microscopes) and instead maintained a single code for ``Lab Equipment.'' Observational studies similarly track students' use of lab equipment without documenting which equipment is being used \cite{dew2024group,quinn2020group,dew2022so}. Once this code book was completed, we imported it into a Jupyter notebook for network analysis.

\begin{figure*}
    \centering
    \includegraphics[width=1\linewidth]{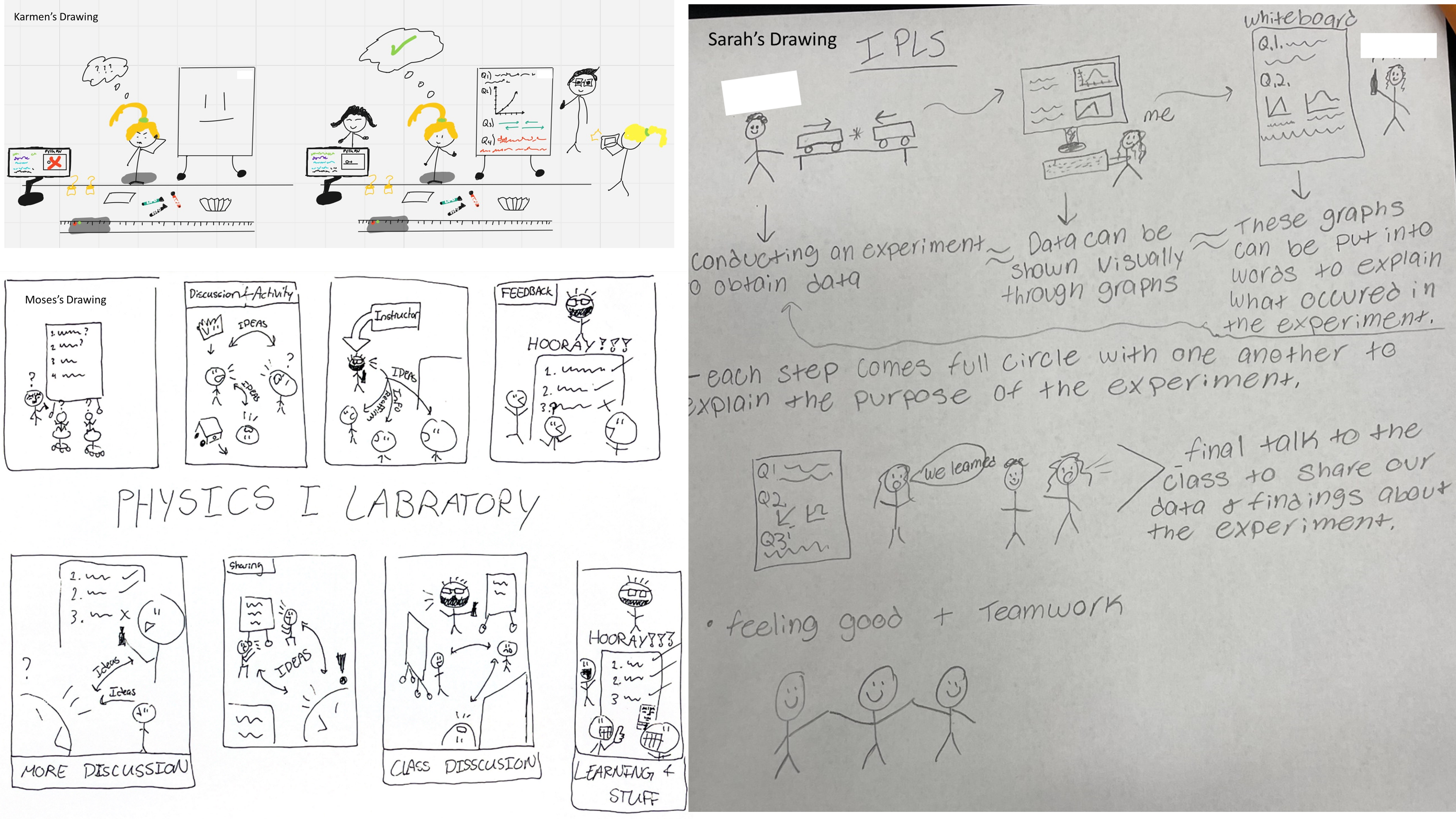}
    \caption{Sample drawings from three lab partners, pseudonyms Karmen, Moses, and Sarah. These pseudonyms were chosen to reflect each student's gender. We discuss examples from the element cataloging process for each of these drawings in Section \ref{subsubsec:ilds}.}
    \label{fig:sample-drawings}
\end{figure*}

As an example of this cataloging process, here we share a few elements that were identified in drawings created by three students from the same lab group. Figure \ref{fig:sample-drawings} shows the drawings submitted by lab partners with pseudonyms Karmen, Moses, and Sarah, and Table \ref{tab:catalog} lists some of the elements we cataloged in each drawing with relevant visual snippets.

\begin{table*}
    \centering
    \begin{tabular}{|m{0.20\textwidth}|m{0.25\textwidth}|m{0.51\textwidth}|}
    \hline \multicolumn{1}{|c|}{Element} & \multicolumn{1}{c|}{Description} & \multicolumn{1}{c|}{Example Snippets} \\ \hline  
    Positive Feelings & The drawing includes some expression of happiness, enjoyment, or satisfaction. Cataloging this element requires more than a smiling face such as a thumbs up, high five, or exclamation. & \includegraphics[width=0.19\textwidth]{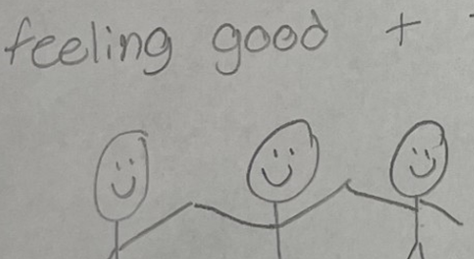} \includegraphics[width=0.09\textwidth]{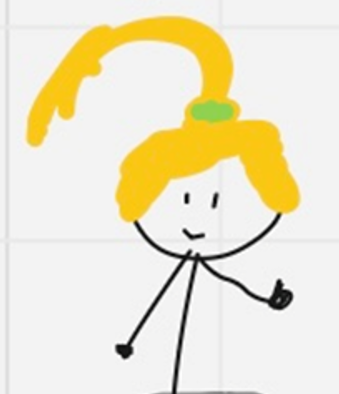} \includegraphics[width=0.19\textwidth]{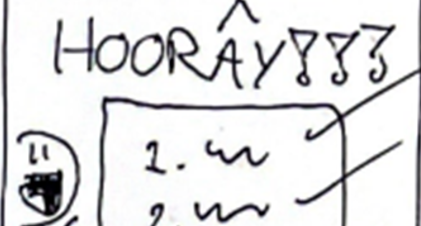} \\ \hline  
    Computer & The drawing includes a computer (desktop, laptop, or tablet) used in the studio activity. & \includegraphics[width=0.19\textwidth]{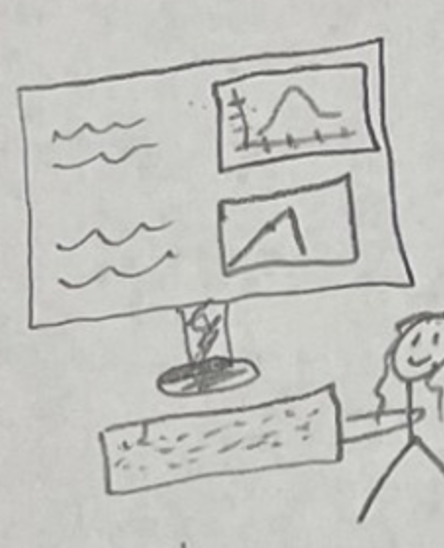} \includegraphics[width=0.19\textwidth]{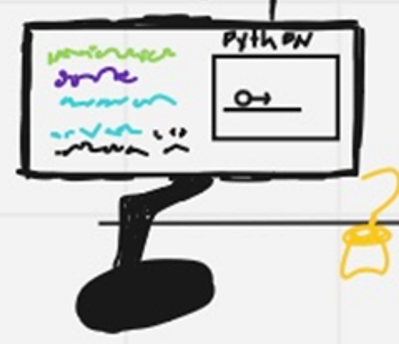} \\ \hline  
    Software & The drawing includes a recognizable or named software (usually data collection, spreadsheet, video tracking, or VPython) used in the studio activity. & \includegraphics[width=0.19\textwidth]{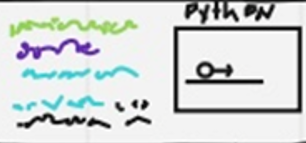} \\ \hline  
    % Ideas & The drawing depicts or labels ideas the students are developing, writing, or sharing. & \includegraphics[width=0.19\textwidth]{ideas-1.png} \\ \hline  
    Discussion & The drawing depicts a back-and-forth conversation between people. There must be an indication of dialogue and not just speaking. & \includegraphics[width=0.19\textwidth]{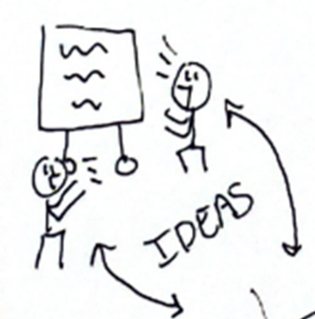} \\ \hline    
        \end{tabular}
    \caption{Sample catalog of some drawing elements from Figure \ref{fig:sample-drawings}.} 
    \label{tab:catalog}
\end{table*}

We coded all three drawings with ``Positive Feelings,'' since Karmen drew a contrast between frowning and smiling, Moses drew exaggerated smiles and a ``Hooray'' exclamation, and Sarah drew high fives with the caption ``feeling good.'' Sarah's drawing explicitly identifies herself and her lab partners by name, so we coded this drawing with ``Group Members'' and ``Self.'' Karmen and Moses' drawings do not contain such labels, but their written descriptions specify that the stick figures represent themselves and their lab partners, so we also coded each of these drawings as depicting ``Group Members'' and ``Self.'' In cases where the figures were not clearly identified, we instead used a generic ``Students'' code. Karmen and Moses both depict the instructor, so their drawings were coded with ``Instructor.'' Karmen and Sarah both depict a computer, so these were both coded with ``Computer.'' However, only Karmen's drawing depicts enough detail to identify the computer screen's contents as a VPython code (also named in her written description), so we also coded Karmen's drawing with ``Software.'' Moses' drawing centers around the process of sharing ideas, so we coded his picture with ``Discussion'' and ``Ideas.'' 

We were successfully able to categorize all cataloged drawing elements as related to goals, members, or practices within the COP framework. Even mundane items like ``Furniture'' were depicted in relation to the community's structure (e.g., the furniture holds the computer and lab equipment), and its presence or absence could represent an important choice the student made about what to include or omit from their representation.

\subsection{\label{sec:NA}Network Analysis of Drawing Elements}

The new methodology that we offer in this paper is a process of analyzing many student-generated drawings using network analysis. In our network analysis, we represent each drawing element as a \textit{node}, with \textit{edges} connecting each pair of elements that occur in the same drawing. In this subsection, we explain our rationale for using network analysis and summarize our process. The network analysis code that we developed is available for use at \cite{git}, along with our anonymized code book as a sample use case.

\subsubsection{Network Analysis: Justification and Overview}

Network analysis methods allow the researcher to consider a set of student data holistically without reducing that data to a single average numerical value, or a singular shift in scores over the course of instruction. As a drawing-based survey promotes student expression in the collection of data, we find that network-based methods help preserve these expressions in data analysis. 

Particularly, we choose to employ network analysis because we are interested not just in what elements are present in the students' drawings but how those elements relate to each other. Cataloging the presence or absence of elements does not give a holistic picture of how those elements relate to each other, and pairwise correlations are insufficient for studying larger-scale connections between the ideas that students are expressing.

For example, suppose multiple students in the class draw a representation of getting help (practice) from the instructor (member) about particular software (practice) on the computer (practice) to answer an assignment question (goal). These are five elements that might co-occur together in multiple drawings, a pattern that would not be revealed by element frequency or pairwise correlations. Additionally, a student who draws this scene might omit one of these elements: They might draw the software's logo but not the physical computer, or draw the computer but not name the software, or show the instructor inspecting the computer without explicitly mentioning the act of helping. Even if one of these elements is missing from the drawing, it might still be present in the student's mental model as part of this larger idea. 

We need network analysis to help us identify high-frequency conceptual connections between elements \cite{riihiluoma2024network} depicted across a set of drawings. Looking for high-frequency connections means identifying elements that not only occur frequently across the drawings but also mediate connections between other elements. For a given network, we quantify this search by examining each element's \textit{frequency} (How many students drew this element?) and its \textit{centrality} (How strongly is this element connected to others across the network?).

We developed our analytical process by reflecting on a set of guiding questions posed by Traxler \textit{et al}. \cite{traxler2024person}. Although these questions were originally framed regarding social networks (which represent connections between people), we find them applicable to our study, since our network represents perceptions of the social construct of the COP. We answer these questions here to help frame our analytical process.

\begin{enumerate}
    \item \textit{What interests you about the network?} %Traxler \textit{et al}.'s suggested answers including structure (e.g., pattern or size), function (e.g., type of exchanges or resources), strength (e.g., relative durability or variety in bonds), and content (e.g., attitudes and beliefs). 
    We are interested in the structure of the interconnectivity of drawing elements and in the content of how that interconnectivity differs among identity groups.
    \item \textit{Why do you think the network is important?} %Operating from a focus on social networks, Traxler \textit{et al}. suggest answers of social support (important to well-being), social capital (important to power and brokerage), and social influence (important to spreading information and attitudes). 
    We think the network of drawing elements is important because it highlights themes in student perceptions of the learning community. These perceptions impact student engagement and persistence (important to well-being) and differences between identity groups highlight inequities (important to power). Studying these differences can help us formulate best practices for engagement (important to spreading attitudes). In keeping with the charge from Kanim and Cid \cite{kanim2020demographics}, using network analysis of qualitative data in this way allows us to explore differences in experiences between identity groups without attempting to rank those groups or their approaches to learning physics.
    \item \textit{How do you intend to study the network?} We use a mixed-methods approach, which Traxler \textit{et al}. define as any network analysis study ``drawing from both qualitative and quantitative data or using qualitative and quantitative methods of analysis.'' The data we collect is qualitative open expression. We code this data thematically to enable quantitative visualization and analysis. We use a whole network approach, but we consider identity groups by filtering results by demographics, which highlights person-centered traits. We are interested in studying which drawing elements are most central in the collection of ideas represented by the drawings across each identity group.
\end{enumerate}

Additionally, throughout our analytical process, we reminded ourselves that network comparison is an open problem \cite{tantardini2019comparing,riihiluoma2024network,evans2022linking}. There is no singularly correct or superior method, and new algorithms and metrics continue to be explored in the field. We therefore focus on principles and approaches that seem most appropriate to our data set. The following observations of our network have guided these choices:

\begin{itemize}
    \item Since our network is made from free expression, there is no definitive list of nodes we expect to observe, only the nodes we have observed. This means that the subnetworks we create from identity groups are not guaranteed to share the same set of nodes. This observation situates our work differently than network analysis studies of multiple-choice instruments \cite{riihiluoma2024network,dalka2024network}.
    \item Our network is relatively \textit{sparse}. Its density (the fraction of connections present out of the total number of connections that could exist) is only about 26\%. 
    \item Our network includes many lower-frequency nodes that may not produce a sufficient level of significance under analysis. As discussed below, this prompts us to prune the lowest-frequency nodes from our network.
    \item The layout of our network (Figure \ref{fig:network}) features a single central cluster of prominent elements surrounded by lower-frequency elements around the periphery. Following examples of clustering techniques in the PER literature \cite{speirs2024utilizing,riihiluoma2024network,dalka2024network}, we found no significant clustering patterns or significant differences in clustering between identity groups. Therefore, we do not discuss node clusters in our results.
\end{itemize}

\subsubsection{Network Representation and Comparison Metrics}

As depicted in Figure \ref{fig:network}, we represent each drawing element as a node with a diameter proportional to the number of drawings this element was found in. Each node is color-coded based on its category within the COP framework (goal, member, practice). For each pair of elements that occur in the same drawing, we connect their nodes with an edge. An edge's weight is proportional to the number of drawings these two elements were found in together. Visually depicting this network immediately gives a general idea of which nodes occur most frequently and hold the most central positions within the network.

We create our networks using the Pandas and Networkx libraries for Python in a Jupyter notebook. We use Pandas to import our code book and create an \textit{adjacency matrix} for a given network $\alpha$, where each $a_{ij}^{\alpha}$ is the number of drawings in which element $i$ and and element $j$ were found. Note that in our adjacency matrix, $a_{ij}^{\alpha} = a_{ji}^{\alpha}$, making this an undirected network. We use Networkx to create the network diagram and compute network comparison metrics.

We are particularly interested in comparing \textit{subnetworks} within our data set: the set of drawings created by different student identity groups based on gender, college generation, and race. Comparing these subnetworks could offer insight into how these identity groups experience introductory physics lab differently. We create these subnetworks by filtering our data in Pandas by student demographics and then passing the filtered adjacency matrix to Networkx. 

We use node frequency and node betweenness to study our network and its subnetworks. Frequency $f_{i}^\alpha$ is defined as the number of drawings in network $\alpha$ that included element $i$. Similar to \cite{riihiluoma2024network}, we chose to count each element's occurrence only once per drawing, rather than document the number of times a drawing depicts a given element. We report node frequency as a percentage of the total number of students represented in that subnetwork.

While node frequency provides a straightforward comparison between subnetworks, it does not speak to a node's role within each subnetwork, and there is no means of establishing an uncertainty value around frequency to determine effect sizes or significance. To achieve these features, we must consider a node's centrality within the network. Centrality quantifies one node's connections to other nodes. A more central node in a network indicates an element of the students' experience that they connect to many other elements. If, for example, ``collaboration'' was a very central node in a network, it would prompt us to consider who the students see themselves collaborating with, what tasks they are collaborating over, and toward what ends. We find it most helpful to evaluate centrality using each node's betweenness value.

\textit{Betweenness} $b_{i}^{\alpha}$ measures node $i$'s centrality by quantifying how strongly it mediates connections between other nodes in network $\alpha$. A node has high betweenness if it is found along the shortest connection between many pairs of other nodes, which aligns with our goal of finding high-frequency conceptual connections between drawing elements \cite{riihiluoma2024network}. Expressed as a number between 0 and 1, betweenness is a fractional count of how many times node $i$ is found along a geodesic (shortest path) linking two other nodes in network $\alpha$:

\begin{equation}
    b_{i}^{\alpha} = \frac{\sum_{j \neq k \neq i} g_{jk}^{\alpha i}}{\sum_{j \neq k} g_{jk}^{\alpha}},
\end{equation}

where $g_{jk}^{\alpha}$ is the number of geodesics between nodes $j$ and $k$ in network $\alpha$ and $g_{jk}^{\alpha i}$ is the number of geodesics between nodes $j$ and $k$ that pass through node $i$. In a network with weighted edges like ours, path lengths are computed using the inverse of edge weight as distance. We find betweenness an appropriate metric to consider in our network, since it quantifies how directly a node is connected with the rest of the network, and how often it serves as a mediator between two other nodes. This conception of centrality aligns with our earlier discussion about elements occurring together to communicate an overall idea. 

For readers already familiar with network analysis, we explain in the Appendix why we omit other centrality measures (closeness, degree, and strength) from this discussion.

% \textbf{Within-category comparisons}. Our network includes a priori classifications of nodes based on the COP framework (goals, members, practices). To examine differences in how these categories were depicted by student groups, we define the following quantities.

% \textbf{Average intra-category weight}. We define the average weight $\bar{w}^{\alpha c}$ within a category $c$ (goals, members, practices) as the sum of all edge weights between nodes in the category divided by the total possible weight:

% \begin{equation}
%     \bar{w}^{\alpha c} = \frac{\sum_{j \neq i \in c} a_{ij}^{\alpha}}{(N_{n}^{\alpha}-1)(N_{n}^{\alpha}-2)N_{d}^{\alpha}},
% \end{equation}

% where $N_{d}^{\alpha}$ is the number of drawings in network $\alpha$. This average weight takes on a value between 0 and 1 and indicates how strongly interconnected category $c$ is within the network.

% \textbf{Average category betweenness}. We also calculate the average betweenness $\bar{b}^{\alpha c}$ of all nodes within category $c$ as a measure of how central the category as a whole is to the network. However, the presence of many small peripheral nodes, such as our network has, could suppress the influence of large central nodes. Therefore, we use a generalized mean with power 2:

% \begin{equation}
%     \bar{b}^{\alpha c} = \left(\frac{1}{N^{\alpha c}}\sum_{i \in c} \left(b_{ij}^{\alpha}\right)^2\right)^{1/2},
% \end{equation}

% where $N^{\alpha c}$ is the number of nodes in category $c$.

\subsubsection{\label{sec:bootstrap_method}Bootstrapping and Effect Sizes}

One desired outcome of this methodology is to find significant differences in node betweenness values between subnetworks of identity groups. Finding significant differences between groups would help answer our research questions by demonstrating a way to analyze qualitative data from drawn expressions, and by demonstrating the types of insights network analysis can show about students' perceptions of the introductory lab as a COP. This process requires comparing the difference between two betweenness values $b_{i}^{\alpha}-b_{i}^{\beta}$ from subnetworks $\alpha,\beta$ against the statistical uncertainties $\sigma_{b_{i}^{\alpha}},\sigma_{b_{i}^{\beta}}$ in those betweenness values. However, we cannot obtain such an uncertainty value experimentally, which would require running multiple instances of IPLS to obtain statistically independent sets of drawings. 

This is a common conundrum in network analysis. For example, suppose one wanted to study networks of flights between airports across the United States, treating each airport as a node and flights as edges and calculating the betweenness of each airport over, say, a year of flights \cite{rosvall2010mapping}. One might pose the question of whether the network changes significantly from one year to the next, which would require comparing changes in $b_{i}^{\alpha}$ for each airport $i$ between two years $\alpha$ against a statistical uncertainty in $b_{i}^{\alpha}$. The conundrum is that one cannot experimentally sample different iterations of US airport arrangements.

\textit{Bootstrapping} is a process that allows us to simulate the effect of random variations within a network using existing network data. In this process, we first create many bootstrap networks as random perturbations of an existing network $\alpha$. Following the example in \cite{rosvall2010mapping}, we create bootstrap networks by randomizing each edge weight $a_{ij}^{\alpha}$ in the adjacency matrix using a Poisson distribution with the measured edge weight as the average value, treating the network's edges as independent random events. For each bootstrap network, we evaluate the metrics of interest (in this case, betweenness). We then evaluate the average and standard deviation for those metrics across all bootstrap networks. With a sufficient number of bootstrap networks (1200, as we establish below), we can test for significant differences between subnetworks.

When comparing two subnetworks $\alpha, \beta$, we determine significance using $p$ values and effect sizes using Cohen's $d$:

\begin{equation}
    d_i^{\alpha,\beta} = \frac{\left|b_{i}^{\alpha}-b_{i}^{\beta}\right|}{\sigma_{b_{i}}^{\alpha,\beta}}.
\end{equation}

Here, $\sigma_{b_{i}}^{\alpha,\beta}$ is the pooled standard deviation of the betweenness values obtained from bootstrapping subnetworks $\alpha$ and $\beta$ for element $i$:

\begin{equation}
       \sigma_{b_{i}}^{\alpha,\beta} =  \sqrt{\frac{\left(N_{d}^{\alpha}-1\right) \left(\sigma_{b_{i}^{\alpha}}\right)^2+\left(N_{d}^{\beta}-1\right) \left(\sigma_{b_{i}^{\beta}}\right)^2}{N_{d}^{\alpha}+N_{d}^{\beta}-2}},
\end{equation}

where $N_{d}^{\alpha}$ is the number of drawings in subnetwork $\alpha$. Following standard practices, $0.5\leq d< 0.8$ is considered a medium effect size (marked in our tables with a *) and $d\geq 0.8$ is considered a large effect size (marked with a **). While we denote medium effect sizes in our results, we limit our discussion to differences between subnetworks that show large effect sizes since these effect sizes can vary with network composition (see Section \ref{sec:limitations}).

Some readers may be familiar with the draw with replacement procedure used in previous network analysis studies in PER \cite{speirs2024utilizing,dalka2024network,riihiluoma2024network}, which is an alternative to the edge weight randomization used here. We note that each of these studies collected fixed-response data from multiple-choice instruments, resulting in a consistent set of nodes across each student's response. In contrast, our drawing-based survey is open-ended, and not every node appears in each student's response. When we investigated using drawing with replacement to bootstrap our network, we found it caused nodes to disappear between different bootstrap iterations, which produced unreliable results.

\section{\label{sec:authors}Author Information and Instructional Context}

The authors of this article include a faculty member (WBL) with 18 years of experience in higher education, a first-year undergraduate physics major (TS), a fourth-year undergraduate neuroscience major pursuing medical school (GK), and a post-baccalaureate psychology graduate pursuing medical school (CD). All three students are female and one is a member of an underrepresented group while the faculty member is white and male. We found this diversity of academic and demographic identities to be crucial in the development of our survey and discussion of the drawings by helping us connect the student experience with the features of the COP framework. WBL and GK adapted the introductory lab drawing survey from a previous version about research group experiences published in \cite{lane2024using}, and validated the new version for use in the introductory lab context using the response process evaluation method \cite{wolf2022survey}. All authors categorized drawing elements and contributed to the network analysis.

We are members of a relatively new PER group at a mid-size regional state university with R2 status. Our university's student body features 60\% female students and 33\% racial minorities who are underrepresented in physics. Our physics department strongly supports engaged-learning practices. We collected responses to the introductory lab drawing survey from students in a two-semester studio-format IPLS sequence. The curriculum for this IPLS sequence follows Michigan State University's Physics at the Molecular and Cellular Level \cite{lapidus2020physics}, which focuses on physics content and experimental skills relevant to life science majors. Both courses in the sequence were formatted similarly, with the instructional team meeting regularly throughout the semester to coordinate delivery. The instructional team was composed of three male and one non-binary faculty members. Student enrollment in these courses was distributed across multiple studio sections meeting at different times throughout the week.

Studio activities in these courses were designed with varying levels of structured inquiry or guided inquiry, depending on the students' progress within the semester or within each unit. Activities incorporated a breadth of mathematical and computational modeling, data collection and analysis, algebraic problem-solving, and multiple representations. New concepts were developed within the studio sessions and then summarized and applied to problem-solving in a large-group lecture session. The sequence of physics topics primarily followed traditional emphases on forces and energy with applications to life science contexts, but replaced rotational motion with a unit on statistical mechanics more relevant to the molecular and cellular context.

\section{\label{sec:results}Results}

Here, we discuss the results of our network analysis of drawing elements depicted by students from a two-semester IPLS sequence. We noticed no significant differences in the type of content drawn by students in the two courses or the degree of detail. Here, we focus our discussion on comparing subnetworks of drawings created by female students (F) and male students (M), drawings created by continuing-generation students (CG) and first-generation students (FG), and drawings created by Asian (A), Hispanic (H), and White (W) students. The drawing element frequencies and betweenness values for each subnetwork are shown in Tables \ref{tab:centralities} and \ref{tab:racial}. 

We used the bootstrapping procedure to determine the standard deviations reported in parentheses and the $p$ values reported in the table captions. All subnetwork comparisons produced $p$ values less than $.001$. We used the standard deviations to determine effect sizes between subnetworks. Medium effect sizes are denoted with a single asterisk and large effect sizes are denoted with a double asterisk. Elements that have a large effect size are further highlighted by bolding and underlining the highest betweenness value. We interpret a large effect size for element $i$ as indicating a difference in the centrality that element $i$ holds within the subnetworks being compared. 

For simplicity, we include in our analysis only the elements that occurred in 5 or more students' drawings. Including nodes with frequency of 4 or less increases a few of the effect sizes reported in this paper, since having more peripheral nodes creates more geodesics that the central nodes lie along. We think it is methodologically safer to underestimate rather than overestimate effect sizes, and can investigate the centrality of these omitted nodes with a larger sample size in a future study.

\subsection{Whole Network Analysis}

\begin{figure*}
    \centering
    \includegraphics[width=1\linewidth]{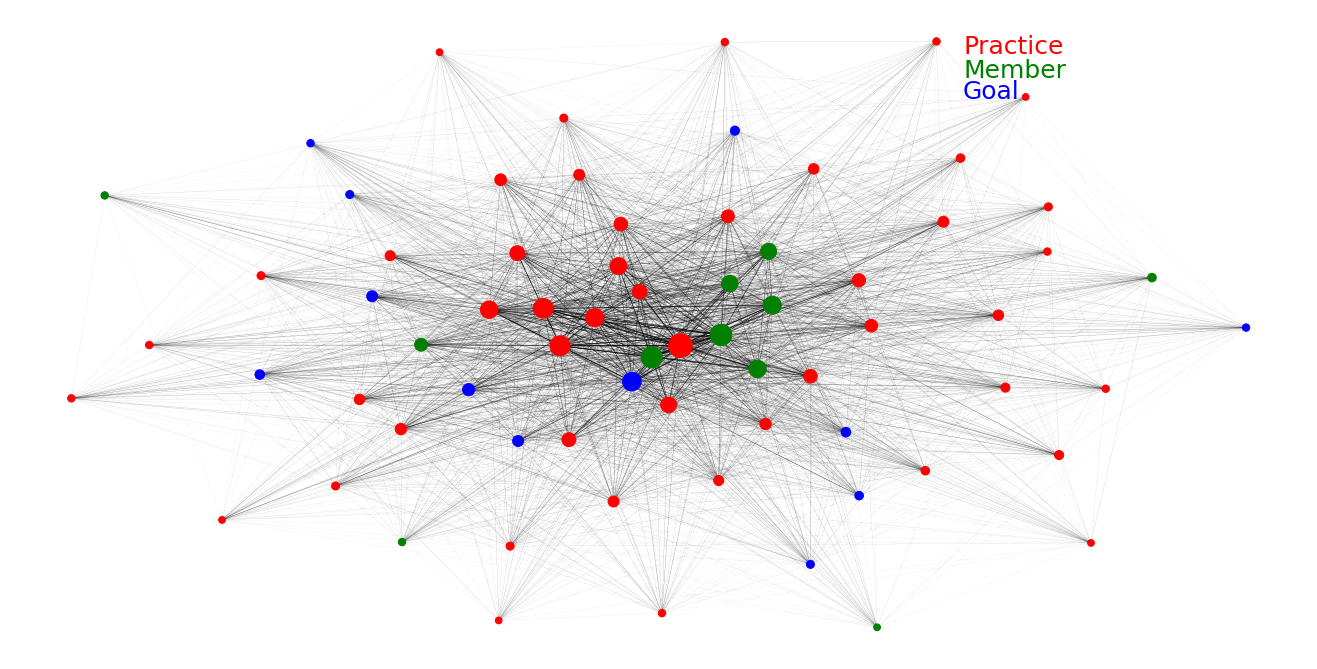}
    \caption{Network diagram for the $N = 74$ student drawings. Each node represents a drawing element, with diameter proportional to element frequency and color representing the element's category from the Communities of Practice framework. Elements that were depicted in fewer than 5 drawings are omitted. Each edge represents two elements occurring in the same drawing, with thickness representing the number of co-occurrences.}
    \label{fig:network}
\end{figure*}

Students depicted an average of 18.7(8.1) elements in their drawings, with the lowest-detail drawing having 3 elements and the highest-detail having 39 elements. Figure \ref{fig:network} shows the network of drawing elements created by all $N = 74$ students, with a total of $71$ cataloged elements. Again, each node represents a single element with diameter corresponding to the element's frequency $f^{\alpha}_{i}$ across the set of drawings and color indicating the node's category in the COP framework. Nodes are connected by edges if they appear in the same drawing, with edge thickness corresponding to the number of co-occurrences $a^{\alpha}_{ij}$.

The network shows a single central cluster of large, highly-connected nodes that represent the core of the students' expressions, with a diversity of smaller nodes around the periphery. Most nodes in the center, and throughout the network, are categorized as practice-related, with the student drawings depicting an average of $11.9(6.1)$ practice-related elements and a total of $47$ practice-related elements in the network. This is unsurprising, as the studio lab is an active learning environment featuring a diversity of tasks, physical objects, and representations of physics concepts. Member-related elements form the second most common category, including depictions of people in the studio and their descriptors (e.g., ``Positive Feelings,'' ``Confused,'' ``Helpful''). The student drawings depict an average of $4.45(1.9)$ member-related elements and a total of $11$ member-related elements in the network. Goal-related elements are fewer in number and more sparsely connected throughout the network. The student drawings depicted an average of $2.3(2.0)$ goal-related elements and a total of $13$ goal-related elements in the network. Students' drawings tended to depict fewer goals, without a strong sense of common goals across the network. 

We choose to focus discussion on the 20 highest-frequency nodes in the full network. This set features insightful differences between the identity groups we consider, and comprise nodes depicted by at least 25\% of students in the full network. The columns labeled ``all'' in Table \ref{tab:centralities} show the frequency and betweenness for each of these nodes in the full network. 

\begin{table*}
    \centering
    % \begin{sideways}
    \begin{tabular}{|c|c|c|c|c|c|c|c|c|c|c|c|c|}
%%% This is where you copy-paste the LaTeX-formatted output from Python:
% Drawing &  $f_{i}^{\alpha}$ & $b_{i}^{\alpha}$
% &\multicolumn{2}{c|}{$f_{i}^{\alpha}$ }& \multicolumn{3}{c|}{$b_{i}^{\alpha}$} & \multicolumn{2}{c|}{$f_{i}^{\alpha}$ }& \multicolumn{3}{c|}{$b_{i}^{\alpha}$}\\
% Element (Category) &  All & All 
% &Female & Male & Female & Male & $d$   & CG & FG & CG & FG &$d$ 
% \\
Element (Category) & $f_{i}^{\textrm{all}}$ & $b_{i}^{\textrm{all}}$ & $f_{i}^{\textrm{F}}$ & $f_{i}^{\textrm{M}}$ & $b_{i}^{\textrm{F}}$ & $b_{i}^{\textrm{M}}$ & $d_{i}^{\textrm{F,M}}$ & $f_{i}^{\textrm{CG}}$ & $f_{i}^{\textrm{FG}}$ & $b_{i}^{\textrm{CG}}$ & $b_{i}^{\textrm{FG}}$ & $d_{i}^{\textrm{CG,FG}}$ \\
\hline
Whiteboard (Practice) &  $82.4$ & $.399(53)$ 
&$79.5 $ & $85.7$ & $.330(57)$ & $.376(65) $ & $0.77^{*}$  & $81.8 $ & $83.3$ & $.352(57)$ & $.310(65) $ &$0.71^{*}$
\\
Group Members (Member) &  $67.6$ & $.156(43)$ 
&$70.5 $ & $60.7$ & $\textbf{\underline{.208}}(49)$ & $.099(37) $ & $2.44^{**}$  & $67.3 $ & $72.2$ & $.157(44)$ & $.186(55) $ &$0.62^{*}$
\\
Self (Member) &  $66.2$ & $.122(38)$ 
&$68.2 $ & $60.7$ & $\textbf{\underline{.164}}(43)$ & $.096(39) $ & $1.65^{**}$  & $65.5 $ & $72.2$ & $.115(39)$ & $\textbf{\underline{.201}}(55) $ &$2.00^{**}$
\\
Writing (Practice) &  $59.5$ & $.080(32)$ 
&$63.6 $ & $50.0$ & $.109(36)$ & $.083(34) $ & $0.75^{*}$  & $63.6 $ & $44.4$ & $\textbf{\underline{.116}}(38)$ & $.055(27) $ &$1.72^{**}$
\\
Furniture (Practice) &  $56.8$ & $.044(23)$ 
&$52.3 $ & $60.7$ & $.050(24)$ & $\textbf{\underline{.123}}(43) $ & $2.23^{**}$  & $56.4 $ & $55.6$ & $.063(29)$ & $.078(34) $ &$0.49$
\\
Learning (Goal) &  $51.4$ & $.063(28)$ 
&$52.3 $ & $50.0$ & $.083(32)$ & $\textbf{\underline{.122}}(42) $ & $1.08^{**}$  & $54.5 $ & $44.4$ & $\textbf{\underline{.141}}(43)$ & $.045(24) $ &$2.43^{**}$
\\
Computer (Practice) &  $48.6$ & $.040(21)$ 
&$52.3 $ & $39.3$ & $.067(27)$ & $.054(28) $ & $0.47$  & $56.4 $ & $27.8$ & $\textbf{\underline{.101}}(35)$ & $.020(16) $ &$2.57^{**}$
\\
Instructor (Member) &  $47.3$ & $.024(17)$ 
&$40.9 $ & $53.6$ & $.024(18)$ & $\textbf{\underline{.102}}(37) $ & $2.88^{**}$  & $43.6 $ & $61.1$ & $.038(22)$ & $\textbf{\underline{.077}}(34) $ &$1.51^{**}$
\\
Positive Feelings (Member) &  $44.6$ & $.014(12)$ 
&$50.0 $ & $39.3$ & $.044(23)$ & $.039(23) $ & $0.20$  & $41.8 $ & $50.0$ & $.030(19)$ & $.045(26) $ &$0.70^{*}$
\\
Doing Math (Practice) &  $44.6$ & $.017(15)$ 
&$43.2 $ & $46.4$ & $.027(19)$ & $\textbf{\underline{.059}}(29) $ & $1.33^{**}$  & $49.1 $ & $33.3$ & $\textbf{\underline{.062}}(29)$ & $0.017(16) $ &$1.71^{**}$
\\
Lab Equipment (Practice) &  $41.9$ & $.016(14)$ 
&$40.9 $ & $46.4$ & $.045(23)$ & $.034(23) $ & $0.50$  & $38.2 $ & $50.0$ & $.029(20)$ & $.047(27) $ &$0.79^{*}$
\\
Students (Member) &  $39.2$ & $.004(6)$ 
&$36.4 $ & $42.9$ & $.012(13)$ & $\textbf{\underline{.050}}(26) $ & $1.99^{**}$  & $43.6 $ & $27.8$ & $\textbf{\underline{.025}}(18)$ & $.006(9) $ &$1.12^{**}$
\\
Confused / Stuck (Member) &  $37.8$ & $.012(11)$ 
&$43.2 $ & $32.1$ & $\textbf{\underline{.049}}(25)$ & $.027(19) $ & $0.95^{**}$  & $36.4 $ & $38.9$ & $.030(20)$ & $\textbf{\underline{.052}}(28) $ &$0.99^{**}$
\\
Collaborate (Practice) &  $36.5$ & $.009(10)$ 
&$40.9 $ & $28.6$ & $\textbf{\underline{.040}}(22)$ & $.021(17) $ & $0.94^{**}$  & $41.8 $ & $22.2$ & $\textbf{\underline{.042}}(24)$ & $.013(14) $ &$1.30^{**}$
\\
Marker (Practice) &  $32.4$ & $.004(6)$ 
&$34.1 $ & $28.6$ & $.012(12)$ & $\textbf{\underline{.027}}(20) $ & $0.98^{**}$  & $34.5 $ & $22.2$ & $.017(15)$ & $.007(9) $ &$0.72^{*}$
\\
Questions (Practice) &  $31.1$ & $.007(7)$ 
&$31.8 $ & $28.6$ & $.020(15)$ & $.022(17) $ & $0.11$  & $32.7 $ & $27.8$ & $.021(16)$ & $.015(13) $ &$0.43$
\\
Talking (Practice) &  $28.4$ & $.001(3)$ 
&$31.8 $ & $21.4$ & $.013(13)$ & $.005(9) $ & $0.64^{*}$  & $32.7 $ & $16.7$ & $.010(12)$ & $.005(8) $ &$0.44$
\\
Discussion (Practice) &  $27.0$ & $.002(4)$ 
&$27.3 $ & $25.0$ & $.010(11)$ & $.012(13) $ & $0.15$  & $29.1 $ & $22.2$ & $.012(13)$ & $.007(10) $ &$0.40$
\\
Graphs (Practice) &  $27.0$ & $.001(2)$ 
&$31.8 $ & $17.9$ & $.012(12)$ & $.005(8) $ & $0.60^{*}$  & $25.5 $ & $33.3$ & $.006(10)$ & $\textbf{\underline{.025}}(18) $ &$1.54^{**}$
\\
Software (Practice) &  $25.7$ & $.001(2)$ &$27.3 $ & $25.0$ & $.006(9)$ & $.015(15) $ & $0.79^{*}$  & $23.6 $ & $27.8$ & $.004(8)$ & $\textbf{\underline{.019}}(15) $ &$1.45^{**}$\\

        \end{tabular}
        % \end{sideways}
    \caption{Frequency ($f_{i}^{\alpha}$, as a percentage) and betweenness ($b_{i}^{\alpha}$, as a fraction $0<b<1$) for the 20 most common drawing elements $i$ for all students ($\alpha=$ all, $N=74$), female students ($\alpha=$ F, $N=44$), male students ($\alpha=$ M, $N=28$), continuing-generation students ($\alpha=$ CG, $N=55$), and first-generation students ($\alpha=$ FG, $N=18$). Asterisks on Cohen's $d_{i}^{\alpha,\beta}$ indicate effect size (* medium, ** large) between subnetworks $\alpha,\beta$, with the larger betweenness bolded and underlined. Standard deviations are obtained from the bootstrapping procedure. All $p$ values are $< .001$.}
    \label{tab:centralities}
\end{table*}

Thirteen of these 20 largest nodes are practice-related, including the most central node, ``Whiteboard.'' This element refers to the rolling whiteboards that each lab group used to collaboratively document their work and share their results and learning with other lab groups. We also noted that many students drew these whiteboards at a larger scale than other elements. For example, they might draw the whiteboard towering over the students even though the actual height difference is 1-2 inches. Such larger-than-life scaling is often a visual representation of relative significance \cite{merriman2006using,thomas1998drawing,jolley2001croatian,aronsson1996social}. In addition to holding the largest frequency across the nodes, ``Whiteboard'' has the largest betweenness for the network as a whole and for each subnetwork, indicating the students connected the rolling whiteboards to many other ideas in their drawings.

Another six of the 20 largest nodes are member-related, such as ``Self,'' ``Group Members,'' and ``Instructor.'' When cataloging elements, we found it necessary to distinguish between specifically identified ``Self'' and ``Group Members'' and more generic ``Students,'' as students tended to frame these elements differently in their written descriptions (``My labmates and I are completing a lab...'' versus ``Students are working at the computer...''). The specific identification of the self and the generic depiction of unidentified students might indicate a difference in how students frame their membership in the lab community, and we preserved this difference in how these elements are named in our code book.

Only one goal-related element, ``Learning,'' is included in this set of 20 largest nodes, depicted by slightly more than half of the students. Lower-frequency goals include ``Completing the Assignment,'' ``Getting the Right Answer,'' ``Getting a Good Grade,'' and ``Leaving the Lab.''

\subsection{Bootstrapping Convergence}

\begin{figure}
    \centering
    \includegraphics[width=1\linewidth]{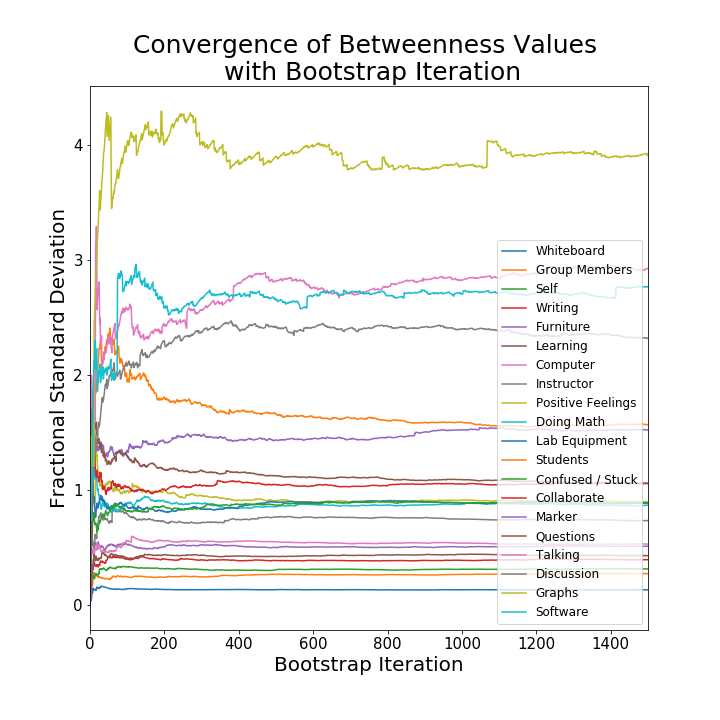}
    \caption{Fractional standard deviation for the betweenness of the 20 highest-frequency drawing elements in the network. Each standard deviation is evaluated for an increasing number of bootstrap iterations, with convergence reached around 1200 bootstraps.}
    \label{fig:bootstrap}
\end{figure}

The Poisson-based bootstrapping process described in Section \ref{sec:bootstrap_method} requires establishing the number of bootstrap networks necessary to converge on a stable uncertainty value. Following the example in \cite{dalka2024network}, we evaluated the standard deviation of each node's betweenness for increasingly large sets of bootstrap networks. We plot these standard deviations versus the number of bootstrap networks in Figure \ref{fig:bootstrap}. Their behavior indicates that the betweenness standard deviations for the 20 highest-frequency nodes converge by about 1200 bootstraps. We therefore use 1200 bootstraps for all betweenness means and standard deviations reported in this paper. This value is similar in magnitude to bootstrap procedures in other PER network analysis studies \cite{speirs2024utilizing,dalka2024network,riihiluoma2024network}. About half of these stabilized standard deviation values are less than the value of the corresponding mean.

% \subsection{Convergence over Sample Size}
%         c. Convergence over Sample Size: Justify sample size

\subsection{Gender Comparison}

Here we compare subnetworks of drawings created by female students ($\alpha=$F, $N = 44$) and male students ($\alpha=$M, $N = 28$). This comparison omits drawings from 2 students from the full network who either reported another gender or did not disclose gender information. Non-gender-conforming students' perspectives of an introductory lab warrant analysis and we plan to collect a focused sample from such students for a future study. 

Female students' drawings featured an average of $19.6(8.5)$ elements, and male students' drawings featured an average of $16.9(7.3)$ elements. We found this difference to be insignificant ($p>.05$), indicating that these identity groups invested indistinguishable amounts of detail in their drawings. We found this pattern across all three element categories (goals, members, practices) from the COP framework. Member-related elements showed the closest-to-significant difference ($p=.07$), which prompts us to consider reexamining these differences with a larger sample size.

Examining the elements in Table \ref{tab:centralities} with large effect sizes based on gender, we observe that female students' drawings feature a central idea of their self and their lab group members (both explicitly identified) collaborating through points of confusion or being stuck. On the other hand, male students' drawings feature a central idea of the instructor interacting with students (not explicitly identified). Female and male students tended to represent learning and doing math with roughly equal frequency, but the male students connected these ideas to more elements, making these elements more central in the male-drawing subnetwork.

\subsection{Continuing/First-Generation Comparison}

Next we compare subnetworks of drawings created by continuing-generation students ($\alpha=$ CG, $N = 55$) and first-generation students ($\alpha=$ FG, $N = 18$). Continuing-generation students' drawings featured an average of $18.6(8.1)$ elements, and first-generation students' drawings featured an average of $18.7(8.2)$ elements. We found this difference to be insignificant ($p>.05$), indicating that these identity groups invested indistinguishable amounts of detail in their drawings. We found this pattern across all three element categories (goals, members, practices) from the COP framework, with no noticeable difference in the frequency with which these students depicted elements from any category.

Examining the elements in Table \ref{tab:centralities} with large effect sizes based on college generation, we observe that continuing-generation students gave more central positioning to the writing process, the goal of learning, the use of computers, the use of mathematics, and collaboration. The first-generation students gave more central positioning to explicitly identified depictions of self instead of drawing generic student figures, although neither subnetwork tended to more strongly centralize their lab group members. The first-generation students also gave more central positioning to the instructor, being confused or stuck, graphs, and software more strongly than the continuing-generation students.

\subsection{Racial Comparison} 

Next we compare subnetworks of drawings created by Asian students ($\alpha=$ A, $N=14$), Hispanic students ($\alpha=$ H, $N=14$), and White students ($\alpha=$ W, $N=44$). This comparison omits drawings from racial groups with three or fewer students represented and students who did not provide information related to racial identity. We plan to collect a focused sample from a greater breadth of racial groups in a future study. This comparison also double counts two students who identified as Asian and White and two students who identified as Hispanic and White. Their responses are included in both subnetworks with which they identify.

Asian students' drawings featured an average of $17.1(7.5)$ elements, Hispanic students' drawings featured an average of $20.8(7.9)$ elements, and White students' drawings featured an average of $20.3(8.4)$ elements. We found these differences to be insignificant ($p>.05$), indicating that these identity groups invested indistinguishable amounts of detail in their drawings. We found this pattern across all three element categories (goals, members, practices) from the COP framework, with no noticeable difference in the frequency with which these students depicted elements from any category.

In Table \ref{tab:racial}, we highlight only elements that produce at least two large effect sizes with a statistically highest betweenness value among all three subnetworks. We find that Asian students tended to depict ``Furniture,'' ``Positive Feelings,'' and generic ``Students'' most centrally among these subnetworks. Hispanic students tended to depict ``Learning,'' ``Lab Equipment,'' and ``Questions'' most centrally among these subnetworks. Finally, White students tended to depict ``Whiteboard,'' ``Group Members'' ``Computer,'' ``Doing Math,'' and ``Collaborate'' most centrally among these subnetworks. Looking at elements that produce at least two large effect sizes with a statistically lowest betweenness value, we see that Asian students depicted ``Group Members'' and ``Self'' the least centrally among these subnetworks, Hispanic students depicted ``Furniture,'' ``Computer,'' and ``Instructor'' the least centrally, and White students depicted ``Graphs'' the least centrally. We found no large effect sizes between these racial groups regarding ``Writing,'' ``Confused / Stuck,'' and ``Talking.'' 

\begin{table*}
    \centering
    \begin{tabular}{|c|c|c|c|c|c|c|c|c|c|}
         Element (Category)  & $f_{i}^{\textrm{A}}$ & $f_{i}^{\textrm{H}}$ & $f_{i}^{\textrm{W}}$ & $b_{i}^{\textrm{A}}$ & $b_{i}^{\textrm{H}}$ & $b_{i}^{\textrm{W}}$ &  $d_{i}^{A,H}$  &  $d_{i}^{H,W}$  &  $d_{i}^{A,W}$ \\
\hline
Whiteboard (Practice)  &  $92.9 $  &  $85.7$  &  $81.8$  &  $.227(54)$  &  $.212(52) $  &  $\textbf{\underline{.357}}(61) $  &  $0.28$ &  $2.49^{**}$ &  $2.08^{**}$\\
Group Members (Member)  &  $57.1 $  &  $78.6$  &  $63.6$  &  $.047(26)$  &  $.163(47) $  &  $\textbf{\underline{.121}}(43) $  &  $3.05^{**}$ &  $0.94^{**}$ &  $1.94^{**}$\\
Self (Member)  &  $57.1 $  &  $71.4$  &  $63.6$  &  $.045(27)$  &  $.099(38) $  &  $.099(41) $  &  $1.61^{**}$ &  $0.03$ &  $1.47^{**}$\\
Writing (Practice)  &  $64.3 $  &  $57.1$  &  $56.8$  &  $.069(32)$  &  $.057(27) $  &  $.066(31) $  &  $0.40$ &  $0.28$ &  $0.10$\\
Furniture (Practice)  &  $78.6 $  &  $57.1$  &  $56.8$  &  $\textbf{\underline{.129}}(43)$  &  $.034(21) $  &  $.063(32) $  &  $2.81^{**}$ &  $1.02^{**}$ &  $1.88^{**}$\\
Learning (Goal)  &  $64.3 $  &  $42.9$  &  $56.8$  &  $.102(38)$  &  $\textbf{\underline{.032}}(21) $  &  $.096(36) $  &  $2.28^{**}$ &  $1.97^{**}$ &  $0.11$\\
Computer (Practice)  &  $42.9 $  &  $35.7$  &  $61.4$  &  $.034(21)$  &  $.016(14) $  &  $\textbf{\underline{.136}}(44) $  &  $1.00^{**}$ &  $3.06^{**}$ &  $2.57^{**}$\\
Instructor (Member)  &  $50.0 $  &  $35.7$  &  $50.0$  &  $.049(27)$  &  $.012(12) $  &  $.046(28) $  &  $1.82^{**}$ &  $1.41^{**}$ &  $0.14$\\
Positive Feelings (Member)  &  $57.1 $  &  $42.9$  &  $45.5$  &  $\textbf{\underline{.063}}(31)$  &  $.015(13) $  &  $.019(17) $  &  $2.03^{**}$ &  $0.22$ &  $2.08^{**}$\\
Doing Math (Practice) &  $35.7 $  &  $28.6$  &  $47.7$  &  $.009(12)$  &  $.009(10) $  &  $\textbf{\underline{.031}}(21) $  &  $0.01$ &  $1.11^{**}$ &  $1.13^{**}$\\
Lab Equipment (Practice)  &  $35.7 $  &  $57.1$  &  $36.4$  &  $.011(12)$  &  $\textbf{\underline{.049}}(26) $  &  $.021(18) $  &  $1.89^{**}$ &  $1.39^{**}$ &  $0.63^{*}$\\
Students (Member)  &  $50.0 $  &  $35.7$  &  $38.6$  &  $\textbf{\underline{.030}}(20)$  &  $.004(6) $  &  $.010(13) $  &  $1.81^{**}$ &  $0.61^{*}$ &  $1.32^{**}$\\
Confused / Stuck (Member)  &  $35.7 $  &  $42.9$  &  $43.2$  &  $.025(18)$  &  $.018(15) $  &  $.028(20) $  &  $0.41$ &  $0.49$ &  $0.17$\\
Collaborate (Practice)  &  $28.6 $  &  $28.6$  &  $40.9$  &  $.006(9)$  &  $.009(9) $  &  $\textbf{\underline{.026}}(21) $  &  $0.32$ &  $0.97^{**}$ &  $1.17^{**}$\\
Marker (Practice)  &  $35.7 $  &  $42.9$  &  $34.1$  &  $.020(17)$  &  $.022(18) $  &  $.011(12) $  &  $0.12$ &  $0.80^{**}$ &  $0.57^{*}$\\
Questions (Practice)  &  $21.4 $  &  $57.1$  &  $31.8$  &  $.007(9)$  &  $\textbf{\underline{.073}}(32) $  &  $.017(15) $  &  $2.78^{**}$ &  $2.76^{**}$ &  $0.67^{*}$\\
Talking (Practice)  &  $21.4 $  &  $21.4$  &  $36.4$  &  $.005(8)$  &  $.003(6) $  &  $.009(11) $  &  $0.36$ &  $0.65^{*}$ &  $0.28$\\
Discussion (Practice)  &  $28.6 $  &  $28.6$  &  $25.0$  &  $.012(13)$  &  $.006(8) $  &  $.004(7) $  &  $0.56^{*}$ &  $0.13$ &  $0.81^{**}$\\
Graphs (Practice)  &  $35.7 $  &  $35.7$  &  $22.7$  &  $.015(15)$  &  $.013(13) $  &  $.001(4) $  &  $0.10$ &  $1.61^{**}$ &  $1.74^{**}$\\
Software (Practice)  &  $21.4 $  &  $35.7$  &  $22.7$  &  $.005(8)$  &  $.014(13) $  &  $.002(5) $  &  $0.79^{*}$ &  $1.61^{**}$ &  $0.58^{*}$\\

    \end{tabular}
    \caption{Frequency ($f_{i}^{\alpha}$, as a percentage) and betweenness ($b_{i}^{\alpha}$, as a fraction $0<b<1$) for the 20 most common drawing elements $i$ for Asian students ($\alpha=$ A, $N=14$), Hispanic students ($\alpha=$ H, $N=14$), and White students ($\alpha=$ W, $N=44$). Our overall network included 3 Black students and 2 Middle Eastern students whose drawings we will include when analyzing a larger sample. For elements that produce at least two large effect sizes with a statistically highest value among the three subnetworks, the largest betweenness value is bold and underlined. All $p$ values are $< .001$.}
    \label{tab:racial}
\end{table*}

\subsection{Gender and Generation Intersections}

The differences between drawings created by different identity groups prompt us to consider how the intersection of these identities influences the centrality of drawing elements. However, we must keep in mind that this further disaggregation leads to some differences in proportionality. Starting with a breakdown by gender, we find that both the female students and the male students retain roughly the same proportion of White students ($59.1\%$ White among female students, $57.1\%$ White among male students, and $59.5\%$ White among the entire sample), with a slightly higher percentage of Hispanic students among female students ($22.7\%$) than among male students ($14.3\%$), and a slightly higher percentage of Asian students among male students ($25.0\%$) than among female students ($15.9\%$). Female students had a slightly higher proportion of first-generation students ($29.5\%$) than the population overall ($24.3\%$), and male students had a slightly higher proportion of continuing-generation students ($82.1\%$) than the population overall ($74.3\%$).

Starting instead with breakdown by generation, among continuing-generation students, the gender breakdown is $54.5\%$ female and $41.8\%$ male, which is similar to the gender breakdown of the sample overall ($59.5\%$ female and $37.8\%$ male). However, among first-generation students, there are proportionally more female students ($72.2\%$ female and $27.8\%$ male). Similarly, among continuing-generation students, the racial breakdown is $16.4\%$ Asian, $16.4\%$ Hispanic, and $67.3\%$ White, similar to the racial breakdown of the sample overall ($18.9\%$ Asian, $18.9\%$ Hispanic, and $59.5\%$ White), but among first-generation students, there are proportionally fewer White students ($33.3\%$ White, $27.8\%$ each of Asian and Hispanic students). Therefore, the results from continuing-generation students might preserve the proportionality of the overall population, but the results from first-generation students might be overly representative of female students and racial minorities. Ironically, this overrepresentation in our data corresponds to an intersection of groups usually underrepresented in PER \cite{kanim2020demographics}. 

We therefore proceed with examining the intersection of gender and college generation since this limits the scale of analysis to only 4 intersection groups in our sample and since the largest overrepresentation (first-generation students having a higher proportion of racial minorities and female students) highlights groups that are typically underrepresented. In Table \ref{tab:subgroups}, we present frequency and betweenness values for the 20 highest-frequency elements drawn by female continuing-generation students (FCG, $N = 30$), female first-generation students (FFG, $N = 13$), male continuing-generation students (MCG, $N = 23$), and male first-generation students (MFG, $N = 5$). We report effect sizes $d_{i}^{\alpha,\beta}$ for each pair of subnetworks $\alpha,\beta$ with a gender or generation trait in common. For example, Column F shows values for $d_{i}^{\textrm{FCG,FFG}}$ (subdividing female students by generation) and Column CG shows values for $d_{i}^{\textrm{FCG,MCG}}$ (subdividing continuing-generation students by gender).

\begin{table*}
    \centering
    \begin{tabular}{|c|c|c|c|c|c|c|c|c|c|c|c|c|}

% Element (Category)  & $f_{i}^{\textrm{FCG}}$ & $f_{i}^{\textrm{FFG}}$ & $f_{i}^{\textrm{MCG}}$ & $f_{i}^{\textrm{MFG}}$ & $b_{i}^{\textrm{FCG}}$ & $b_{i}^{\textrm{FFG}}$ & $b_{i}^{\textrm{MCG}}$ & $b_{i}^{\textrm{MFG}}$ &  $d^{\textrm{FCG,FFG}}_{i}$  &  $d^{\textrm{MCG,MFG}}_{i}$  &  $d^{\textrm{MCG,FCG}}_{i}$  &  $d^{\textrm{MFG,FFG}}_{i}$ \\
 & \multicolumn{4}{c|}{$f_{i}^{\alpha}$} & \multicolumn{4}{c|}{$b_{i}^{\alpha}$} & \multicolumn{4}{c|}{$d_{i}^{\alpha,\beta}$} \\
Element (Category)  & FCG  & FFG & MCG & MFG & FCG & FFG & MCG & MFG &  F & M  & CG  & FG \\
\hline
Whiteboard (Practice)  &  $80.0 $  &  $76.9$  &  $82.6 $  &  $100.0$  &  $.237(59)$  &  $.229(60) $  &  $.296(62)$  &  $.062(21) $  &  $0.13$ &  $4.05^{**}$ &  $1.01^{**}$ &  $2.69^{**}$\\
Group Members (Member)  &  $73.3 $  &  $69.2$  &  $56.5 $  &  $80.0$  & $ \textbf{\underline{.203}}(53)$ &  $.142(52) $  &  $.062(32)$  &  $.036(17) $  &  $1.16^{**}$ &  $0.84^{**}$ &  $3.06^{**}$ &  $2.12^{**}$\\
Self (Member)  &  $66.7 $  &  $76.9$  &  $60.9 $  &  $60.0$  &  $.118(43)$  & $ \textbf{\underline{.192}}(60) $ &  $.076(34)$  &  $.019(12) $  &  $1.50^{**}$ &  $1.81^{**}$ &  $1.03^{**}$ &  $3.18^{**}$\\
Writing (Practice)  &  $70.0 $  &  $46.2$  &  $52.2 $  &  $40.0$  & $ \textbf{\underline{.118}}(44)$  &  $.043(24) $  &  $.070(33)$  &  $.013(10) $  &  $1.91^{**}$ &  $1.90^{**}$ &  $1.09^{**}$ &  $1.16^{**}$\\
Furniture (Practice)  &  $50.0 $  &  $53.8$  &  $60.9 $  &  $60.0$  &  $.035(24)$  &  $.055(30) $  &  $.106(41)$  &  $.023(14) $  &  $0.75^{*}$ &  $2.16^{**}$ &  $2.13^{**}$ &  $0.86^{**}$\\
Learning (Goal)  &  $56.7 $  &  $46.2$  &  $52.2 $  &  $40.0$  &  $.096(40)$  &  $.027(19) $  &  $.105(40)$  &  $.008(8) $  &  $1.98^{**}$ &  $2.61^{**}$ &  $0.13$ &  $0.93^{**}$\\
Computer (Practice)  &  $63.3 $  &  $30.8$  &  $43.5 $  &  $20.0$  & $ \textbf{\underline{.116}}(43)$  &  $.014(14) $  &  $.049(28)$  &  $.003(3) $  &  $2.79^{**}$ &  $1.76^{**}$ &  $1.80^{**}$ &  $0.90^{**}$\\
Instructor (Member)  &  $30.0 $  &  $69.2$  &  $56.5 $  &  $40.0$  &  $.008(12)$  &  $.090(41) $  & $ \textbf{\underline{.115}}(43)$  &  $.004(6) $  &  $3.35^{**}$ &  $2.77^{**}$ &  $3.66^{**}$ &  $2.41^{**}$\\
Positive Feelings (Member)  &  $46.7 $  &  $53.8$  &  $39.1 $  &  $40.0$  &  $.030(23)$  &  $.032(23) $  &  $.036(23)$  &  $.007(7) $  &  $0.09$ &  $1.37^{**}$ &  $0.21$ &  $1.27^{**}$\\
Doing Math (Practice)  &  $50.0 $  &  $30.8$  &  $47.8 $  &  $40.0$  &  $.050(28)$  &  $.017(17) $  &  $.056(30)$  &  $.006(7) $  &  $1.32^{**}$ &  $1.77^{**}$ &  $0.17$ &  $0.60^{*}$\\
Lab Equipment (Practice)  &  $40.0 $  &  $38.5$  &  $39.1 $  &  $80.0$  &  $.033(22)$  &  $.025(21) $  &  $.020(18)$  &  $.027(15) $  &  $0.35$ &  $0.38$ &  $0.63^{*}$ &  $0.46$\\
Students (Member)  &  $40.0 $  &  $30.8$  &  $47.8 $  &  $20.0$  &  $.011(14)$  &  $.003(6) $  &  $.052(28)$  &  $.001(3) $  &  $0.64^{*}$ &  $1.92^{**}$ &  $1.96^{**}$ &  $0.26$\\
Confused / Stuck (Member)  &  $36.7 $  &  $53.8$  &  $39.1 $  &  $0.0$  &  $.015(16)$  & $ \textbf{\underline{.072}}(35) $ &  $.031(21)$  &  $0$  &  $2.46^{**}$ &  $1.60^{**}$ &  $0.97^{**}$ &  $2.33^{**}$\\
Collaborate (Practice)  &  $46.7 $  &  $30.8$  &  $34.8 $  &  $0.0$  & $ \textbf{\underline{.053}}(29)$  &  $.012(13) $  &  $.022(18)$  &  $0$  &  $1.62^{**}$ &  $1.33^{**}$ &  $1.23^{**}$ &  $0.97^{**}$\\
Marker (Practice)  &  $33.3 $  &  $30.8$  &  $34.8 $  &  $0.0$  &  $.011(13)$  &  $.009(10) $  &  $.034(22)$  &  $0$  &  $0.12$ &  $1.65^{**}$ &  $1.36^{**}$ &  $0.99^{**}$\\
Questions (Practice)  &  $30.0 $  &  $38.5$  &  $34.8 $  &  $0.0$  &  $.014(14)$  &  $.034(23) $  &  $.026(19)$  &  $0$  &  $1.15^{**}$ &  $1.47^{**}$ &  $0.77^{*}$ &  $1.70^{**}$\\
Talking (Practice)  &  $40.0 $  &  $15.4$  &  $21.7 $  &  $20.0$  &  $.025(21)$  &  $.001(3) $  &  $.001(4)$  &  $.003(3) $  &  $1.38^{**}$ &  $0.35$ &  $1.48^{**}$ &  $0.54^{*}$\\
Discussion (Practice)  &  $33.3 $  &  $15.4$  &  $21.7 $  &  $40.0$  &  $.023(20)$  &  $.001(2) $  &  $.006(9)$  &  $.010(8) $  &  $1.35^{**}$ &  $0.43$ &  $1.10^{**}$ &  $2.08^{**}$\\
Graphs (Practice)  &  $30.0 $  &  $38.5$  &  $17.4 $  &  $20.0$  &  $.005(9)$  &  $.019(16) $  &  $.002(4)$  &  $.003(3) $  &  $1.25^{**}$ &  $0.25$ &  $0.51^{*}$ &  $1.09^{**}$\\
Software (Practice)  &  $23.3 $  &  $30.8$  &  $26.1 $  &  $20.0$  &  $.004(9)$  &  $.014(13) $  &  $.008(11)$  &  $.003(4) $  &  $0.92^{**}$ &  $0.49$ &  $0.35$ &  $0.84^{**}$\\

    \end{tabular}
    \caption{Frequency ($f_{i}^{\alpha}$, as a percentage) and betweenness ($b_{i}^{\alpha}$, as a fraction $0<b<1$) for the 20 most common drawing elements $i$ for intersections of gender (F, M) and college generation (CG, FG). Four of these elements do not appear in the MFG subnetwork, so those betweenness values are taken as zero. Cohen's $d$ is reported for subnetworks with gender or college generaiton in common: %Column F shows Cohen's $d$ between FCG and FFG, Column M shows Cohen's $d$ between MCG and MFG, Column CG shows Cohen's $d$ between FCG and MCG, and Column FG shows Cohen's $d$ between FFG and MFG. 
    Column F shows $d^{\textrm{FCG,FFG}}_{i}$, Column M shows $d^{\textrm{MCG,MFG}}_{i}$, Column CG shows $d^{\textrm{FCG,MCG}}_{i}$, and Column FG shows $d^{\textrm{FFG,MFG}}_{i}$. For elements that produce large effect sizes for all Cohen's $d$, the largest betweenness value is bold and underlined. All $p$ values are $< .001$.}    
    \label{tab:subgroups}
\end{table*}

Splitting our data in this way greatly reduces the sample sizes we are working with, which could affect the reliability of our effect sizes. For example, four of the 20 largest elements do not appear in the male first-generation students' drawings (our smallest subnetwork), which could produce effect sizes that are arbitrarily large. To accommodate this shortcoming, we restrict our discussion here to elements for which all four effect sizes are large. As before, we bold and underline the largest of the four betweenness values for these elements.

The female subnetworks tended to explicitly identify themselves and their lab group members, while the male subnetworks did not. This suggests that the higher betweenness for ``Self'' in first-generation students (Column $b_{i}^{\textrm{FG}}$ in Table \ref{tab:centralities}) might be attributable to the female students in that subnetwork. On the other hand, the female subnetworks seem split in how centrally they depict writing, computers, and collaboration, with female continuing-generation students and male continuing-generation students both making these elements more central than their first-generation counterparts do. The centrality we saw female students give being confused or stuck seems primarily attributable to female first-generation students, with no male first-generation students depicting this experience. 

We previously noted how male students and first-generation students both tended to centralize the presence of the instructor in their drawings. Here, we see that, among male students, continuing-generation students tended to make their instructor more central, while among female students, first-generation students tended to make their instructor more central. Female continuing-generation students and male first-generation students both decentralized the instructor in their drawings.

Perhaps unsurprisingly, no elements have their highest betweenness among male first-generation students, the subnetwork with the smallest sample size, and the subnetwork missing four of the 20 highest-frequency nodes. Due to the diminishing sample size, we omit further bifurcation of our sample by racial traits, but could carry out this process with a larger sample.

\section{\label{sec:discussion}Discussion}

This proof-of-concept study has demonstrated that we can use network analysis of drawing-based qualitative data to identify differences in identity groups' perceptions of an introductory physics lab. Here, we discuss our findings and how they compare against previous studies of the introductory physics lab experience. First, we revisit the distribution of elements by COP categories (goals, members, and practices). Then, we discuss thematic differences found between identity groups, including approaches to group work, interactions with members of the lab community, and depictions of computers and lab equipment.

\subsection{Element Distribution}

We found that different identity groups depicted an indistinguishable number of elements in their drawings. This similarity reinforces the claim from the literature that drawing-based data collection promotes equity by appealing to students of diverse backgrounds and experiences \cite{picker2000investigating,oistein2023stereotypical,chambers1983stereotypic,hatisaru2022knowledge,cox1999children,chambers1983stereotypic,gameiro2018drawingout,literat2013pencil,merriman2006using}.

Within the full network and each subnetwork, we found different relative frequencies of goal-, member-, and practice-related elements. Visually examining Figure \ref{fig:network} and Tables \ref{tab:centralities} through \ref{tab:subgroups}, we can see that practice-related elements dominate the network. This is perhaps to be expected, as the IPLS studio is a hands-on learning environment with many items of equipment, technical tasks, and ways of collaborating that are new to these students and important for their course outcomes. This prominence of practices illustrates Irving, McPadden, and Caballero's argument that an introductory lab that functions as a COP ``facilitates a focus on practice development'' by engaging students in a shared repertoire of authentic practices \cite{irving2020communities}.

Member-related elements, although fewer in number, feature prominently near the center of the network diagram with only a few toward the periphery. Again, this is unsurprising, as the membership of the studio is clearly established, with collaborative and emotional experiences common across identity groups. We find an interesting contrast in drawings that explicitly identify the student's self and their group members (depicted in two-thirds of the drawings) and drawings that depict generic, unidentified student figures (depicted in just over one-third of the drawings). This distinction is reinforced in how students referred to these figures in their written descriptions (e.g., ``My labmates and I are measuring...'' versus ``Students are measuring...''). We think this contrast between explicit self- and partner-identification is important based on Irving, McPadden, and Caballero's argument that an introductory lab that functions as a COP provides ``space for identity development,'' including a sense of identity within the community, a sense of belonging, and commitment to the community in an environment of respect and trust \cite{irving2020communities}. Identity groups who tend to explicitly depict themselves less centrally than others (in our case, male students and Asian students) might feel less of a sense of identity within the course, possibly indicating disengagement in the learning process. This relationship could be explored by administering the introductory lab drawing survey alongside an instrument to measure identity within the course or STEM identity more broadly. One could then form subnetworks of students with a well-developed or less-developed sense of identity and examine how these students perceive the course differently.

The goal-related elements are located primarily around the network's periphery. ``Learning'' is the only goal among the 20 highest-frequency elements, and is found in just over half the students' drawings. All other goal-related elements occur in fewer than 25\% of the students' drawings. While goal-related elements are more abstract and therefore perhaps more difficult to depict, 18 of the 74 drawings ($24\%$) do not depict any recognizable goal-related elements. The network diagram shows us that, although there are 13 goal-related elements in at least five students' drawings, they are less centrally connected to other elements. This sparsity could indicate a weak consensus among the students about their reasons for taking a physics course, the rationale behind the course's guided inquiry structure, or the expected outcomes of studio activities. Indeed, the goal-related elements hint at a range of student motivations: ``Learning,'' ``Concepts,'' and ``Interesting / Relevant'' likely point to a learning-based motivation while ``Assignment,'' ``Correct / Right,'' and ``Grade / Passing Class'' likely point to a performance- or reward-based motivation \cite{dawson2009learning}. Mirroring an observation from Holmes and Weiman \cite{holmes2018introductory}, 9 of the 71 students depicted a goal of completing the lab activity on time. By comparing this outcome to the ideal of a COP \cite{irving2020communities} and identifying the most commonly held disparate goals among this population of students, this analysis helps us identify clear next steps for course development (discussed below).

% The inconsistency of goal-related elements suggests a next step for course development (discussed below), as an introductory lab that functions as a COP develops a shared set of interests and meaning among students . Holmes and Wieman point out that physicists share a similar inconsistency in goals for introductory labs, with ``goals rang[ing] over reinforcing content, learning about measurement and uncertainty, practicing communication skills, developing teamwork skills, and, more broadly, learning that physics is an experimental science'' \cite{holmes2018introductory}.

\subsection{Differences Between Identity Groups}

Of the 20 highest-frequency drawing elements, we found that 10 exhibited a large effect size when comparing the betweenness values for drawings from female and male students, and that 11 exhibited a large effect size when comparing the betweenness values for drawings from continuing-generation and first-generation students (Table \ref{tab:centralities}). Seven of these large effect sizes persisted when we further disaggregated student drawings by both generation and gender traits (Table \ref{tab:subgroups}), although our sample has a small number of male first-generation students. We find 11 elements with a statistically largest betweenness among Asian, Hispanic, and White students, and six elements with a statistically lowest betweeenness among these groups (Table \ref{tab:racial}). The elements with consistent large effect sizes across gender, college generation, and racial comparisons are ``Group Members,'' ``Self,'' ``Computer,'' ``Instructor,'' and ``Collaborate.'' This consistent difference suggests focusing on these elements in future studies, as it raises instructionally relevant questions like, ``Why do some groups depict the instructor less centrally? Could this lack of depiction indicate a sense of self-sufficiency (student feels well equipped to guide their own learning) or a lack of perceived support (student feels disconnected from the instructor)?''

Several other studies explore gender-based differences in the experience of introductory physics labs. For example, Doucette, Clark, and Singh highlight that ``gendered interactions between students within the masculinized culture of physics may be responsible for students having different learning experiences depending on their gender'' \cite{doucette2022students}. We make the following comparisons between our identity group-based observations and previous literature.

\subsubsection{Approaches to Group Work}

One primary feature of guided inquiry introductory labs is students working together to complete the activity and develop a shared understanding of the content. Dew \textit{et al}. \cite{dew2024group} specify that this ``working together'' could take the form of collaborative learning or cooperative learning. In collaborative learning, each lab group determines how to divide tasks within the learning activity, often leading to a delegation of roles or tasks that each student fulfills in completion of the activity. In cooperative learning, the group work is scaffolded such that ``the goal or outcome is not achievable by individuals working independently and merely collating their work at the end.'' It has been argued that collaborative learning may produce or reveal inequities \cite{shah2019amplifying}, such that guided inquiry labs might not achieve the same improved learning outcomes for all students. Dew \textit{et al}. found that female and male students tended to prefer collaborative and cooperative learning equally \cite{dew2024group}, while Holmes \textit{et al}. \cite{holmes2022evaluating} found that both female and male physics majors preferred cooperative learning. 

Although we did not specifically encourage students to depict their lab group's approach along the collaborative/cooperative learning framework and did not catalog elements explicitly along this framework, we did distinguish in our code book between ``Collaborate'' (a depiction of working together toward a goal, closer to cooperative learning) and ``Roles / Delegation'' (a depiction of group members taking on different tasks, closer to collaborative learning). The ``Collaborate'' and ``Roles / Delegation'' elements are not mutually exclusive, although we identified both elements in only $5.6\%$ of all students' drawings. The ``Collaborate'' element was identified in $36.5\%$ of all students' drawings, with a betweenness of only $.009(10)$. On the other hand, the ``Roles / Delegation'' element was identified in only $9.5\%$ of all students' drawings with a betweenness of $0$. It therefore seems that these students made ``Collaborate'' more central than ``Roles / Delegation,'' possibly pointing to Dew \textit{et al}.'s cooperative learning. We found that female students tended to make ``Collaborate'' more central than male students, and continuing-generation students tended to make ``Collaborate'' more central than first-generation students. This difference persisted when we investigated the intersections of generation and gender, with female continuing-generation students centralizing ``Collaborate'' the  most. We also found that White students tended to make ``Collaborate'' more central than Asian students and more than Hispanic students. These results suggest that the differences in approaches to group work observed by \cite{dew2024group} and \cite{holmes2022evaluating} might be related to how different student groups perceive the introductory lab community.

Within the context of group work where roles are delegated, previous studies have observed female students preferring and taking on a secretary-type role as note-taker or manager \cite{dew2024group,doucette2020hermione}. While our element cataloging process does not record which students are depicted as taking on such a role, we do find that a greater percentage of female students tended to depict writing than male students, although the effect size for the betweenness is not large. We find that a greater percentage of female students than male students drew a student holding a dry erase marker, although this element is more central in the male students' drawings with a large effect size. We also found that continuing-generation students made writing more central than first-generation students, and we found no difference in how central Asian, Hispanic, and White students made writing.

\subsubsection{Interactions with the Studio Community\label{sec:interactions}}

Many introductory labs that follow a guided inquiry approach also encourage interactions between lab groups. These interactions might look like asking questions about experimental setups, comparing data and trendlines, or consulting on answers to activity questions. Sundstrom \textit{et al}. \cite{sundstrom2022examining} used social network analysis to classify lab groups as noninteractors (no observed interactions with other lab groups), information seekers (only engage in intergroup interactions that they initiate), responders (only engage in intergroup interactions that other lab groups initiate), and mutual interactors (engage in both initiated and noninitiated intergroup interactions). They found that ``all-male groups took on interactive roles more often than all-female and mixed-gender groups in... reformed lab sections.'' In our cataloging of drawing elements, we found that $24.3\%$ of student drawings depicted another lab group besides the drawer's, and $14.8\%$ depicted a board meeting in which lab groups presented their work to each other. These elements have a betweenness less than $.001$ in all subnetworks that we study, indicating that no identity groups more strongly centralized interactions with other lab groups, although none made this element very central.

Identity groups in our sample differed in how centrally they depicted an instructor. For example, an instructor features more centrally in male students' drawings than in female students' drawings with a large effect size. Our analysis does not offer an explanation for this difference, but we can think of two plausible explanations: First, we note that three of the four members of the IPLS instructional team this semester were male, with the fourth, non-binary, instructor describing themself as often perceived by students as male. Therefore, it could be that the male students identified more closely with the instructional team, and this identification is reflected in the network of drawing elements. However, we also found that the centrality of an instructor element varied with college generation, with male continuing-generation students and female first-generation students depicting the instructor more centrally than female continuing-generation students and male first-generation students, with a large effect size for all comparisons. This leads us to consider a second explanation, that some students might perceive an instructor as an authority figure, which could play a different role in the perspectives of these intersectional identity groups. These interpretations could be tested with additional survey questions and a coding scheme to specifically investigate the instructors' depictions.

\subsubsection{Lab Equipment and Computers}

Previous studies \cite{dew2024group,holmes2022evaluating} found that male students expressed a preference for using lab equipment slightly more often than female students. Male students have also been observed using the equipment more often \cite{quinn2020group,dew2022so}, although partner agreements seem to lead to more equitable equipment usage \cite{dew2024group}. Our analysis found no large effect sizes for gender-based differences in how students made lab equipment central in their drawings. The effect size for lab equipment between continuing-generation and first-generation students is just below the threshold for large, although this effect size is reduced when we examine the intersection of gender and college generation. We did find a large effect size between Asian students and Hispanic students and between Hispanic students and White students, with Hispanic students making lab equipment more central in both comparisons. 

Studies find that student use of computers by gender can vary, possibly influenced by the purpose of computer usage. Dew \textit{et al}. \cite{dew2024group} found no gendered differences in the use of computers in a lab where students used computers to both analyze data and take notes. However, Day \textit{et al}. \cite{day2016gender} observed male students using computers more than female students in a lab where notes were taken on paper and computers were used only for data analysis. In our study, computers are equally central in the female and male subnetworks. Of note, this IPLS sequence used computers for both analysis and note-taking as in Dew \textit{et al}. \cite{dew2024group}. However, we separately cataloged students' depiction of recognizable named software (e.g., spreadsheets, data collection, VPython), which helps us identify when students assign a specific purpose to the computers. We find that a nearly equal percentage of female and male students depicted named software, with only a medium effect size in betweenness. We also found that computers are more central to the continuing-generation subnetwork than the first-generation subnetwork with a large effect size, but software is more central to the first-generation subnetwork with a large effect size. Computers are also more central to the White subnetwork than the Asian and Hispanic subnetworks with a large effect size. 

These differences seem to indicate that lab equipment and computers feature differently in how these groups think about their introductory lab experience. For many students, these elements represent new and potentially intimidating learning experiences \cite{lunk2016attitudes}, so in both observational studies and this study of student perspectives, it is important that we attend to how students respond to their presence in the introductory lab.

\section{Limitations\label{sec:limitations}}

This proof-of-concept study has demonstrated that we can use network analysis to examine differences in how students express their perceptions of an introductory physics lab in a drawing-based survey. While we cannot claim generalizability of the particular results we found from our sample of IPLS students, this study paves the way for larger-scale investigations. Here, we discuss limitations in this approach, including assumptions behind our use of the survey, social factors that could shape students' responses, and features of our network that limit our study. We also discuss how we have addressed these limitations.

A primary assumption of this study is that students responding to the drawing-based lab survey will choose to depict elements that are important to them. This assumption is supported in the survey's validation study \cite{lane2024using}, but it is always possible that different students respond to a survey differently. To address this concern, we point to the broad diversity of elements and their combinations depicted in Figure \ref{fig:network}. Whether the students' reasoning was conscious or subconscious, they have chosen different elements to depict, and the survey picks up these differences without needing to identify their origin. %In another manuscript \cite{lane2025using} we demonstrate how students in the same lab group created very different depictions, which our element cataloging procedure can detect.

As discussed in Section \ref{sec:WhyDrawing}, while a drawing-based survey can expand students' means of expression, it also potentially places high demand on students' physical and visual abilities, potentially underserving students who experience limitations along those dimensions \cite{scanlon2019ability,mcpadden2023planning}. We are not aware of any such experiences among the student sample discussed here, but our survey's reliance on physical and visual abilities could result in an underrepresentation of student perspectives in the drawing data. The survey does include a prompt for a written description of the student's drawing, which we collected via an on-line survey portal which makes available to students any textual accessibility support (e.g., speech-to-text) that they might be accustomed to using.

Our survey is designed under the assumption that a student's experience in an introductory lab is appropriately modeled as a COP. While this assumption is supported by the literature \cite{irving2014conditions,irving2020communities}, and likely true for many students' experiences, it is not guaranteed that every studio section in our study similarly functions as a COP, or that each lab group or each student similarly experiences the lab as a COP. However, our whole-class network diagram in Figure \ref{fig:network} shows a consistent central experience for our students, with 20 elements being depicted in at least 25\% of the students' drawings, and the remaining elements revealing a diversity of perspectives across this common experience. The fact that the center of the network features so few goals speaks to potential avenues for reinforcing the shared purpose of the course, which we discuss below.

A concern with any survey about students' perceptions of a course is that the students might respond with what the instructor or researcher wishes to see. We reduced this influence by communicating to the students that we would not review their responses until after final grades had been submitted, and scheduling class time for students to complete the survey without their instructors in the room. We can also look at the elements the students depicted to ascertain the likelihood of this influence. We note that less than half of all students depicted an instructor in their drawings. Even fewer depicted a sense of positive feelings about the course, and only $8\%$ of drawings were cataloged as depicting a member of the class community as helpful. When we review the edges between these elements, we find that only $19\%$ of drawings were cataloged with ``Instructor'' and ``Positive Feelings'' and only $4\%$ were cataloged with ``Instructor'' and ``Helpful.'' If the students were responding in such a way as to appease an instructor, we might expect these percentages to be higher.

Applying node pruning and clustering algorithms such as fast-greedy, Louvian, and edge betweenness partitioning did not reveal clusters that were stable under bootstrapping. On the one hand, this network structure points to an overall consistency of student experience. On the other hand, we do find differences in the betweenness of individual elements between subnetworks. With a larger sample size or more appropriate clustering algorithms, these differences could yield stable clusters that highlight thematic differences in student perceptions.

Finally, we noted earlier that we removed smaller nodes (frequency less than 5) from our network for simplicity. Including these nodes increases the difference in betweenness values for some subnetworks, producing larger effect sizes for some nodes than reported here. Such an increase is to be expected, since the presence of more less-connected nodes along the network's periphery would increase the number of geodesics the central nodes lie along. Therefore, including these lower-frequency nodes would exaggerate the centrality of the most central nodes, and we think it safer to underestimate rather than overestimate effect sizes. When we collect surveys from a larger sample, we can reintroduce these lower-frequency nodes as appropriate. Since the effect sizes reported here depend on choices we made about what elements to include in the network, we focused discussion only on large effect sizes among the most central nodes.

\section{\label{sec:implications}Implications for Future Research and Instruction}

The approach demonstrated in this paper has a wide range of uses in future PER studies. Most broadly, this study serves as an example of using network analysis to visualize and quantitatively study qualitative student data. This approach could be adapted to study data from interviews, problem-solving sessions, classroom observations, and written free-response items. The use of network analysis across these types of data collection could help facilitate comparisons between their findings, thereby aiding researchers in using a diverse set of research tools with a diversity of ability demands \cite{scanlon2019ability,mcpadden2023planning}.

The introductory lab drawing survey can also be adapted for investigations using a different framework. For example, the checklist of COP-related items in the survey could be replaced with items related to the collaborative/cooperative learning framework discussed in \cite{dew2024group} to investigate students' perceptions of these approaches to group learning. The process of cataloging elements could also be adjusted to specifically document depictions of these approaches.

The introductory lab drawing survey can be given at different times during the semester to study longitudinal effects. A pre- and post-version could capture how students' perceptions of physics lab change between the first few weeks to the end of the semester. The survey could also be modified to ask incoming students to draw their ideal physics learning environment. Such a projective survey could provide information about their incoming expectations and serve a reflective purpose similar to partner agreements \cite{dew2024group}. 

The drawing survey and network analysis can also be applied to investigate differences in student experiences beyond the context of a single course. Following the example of prior studies \cite{commeford2021characterizing,dew2024group}, this approach can be used to compare student perceptions across different lab formats such as studio, ISLE, or Tutorials in Introductory Physics. The drawing survey can also be used to compare students' perceptions of labs from different subjects. When we collected the survey responses for this paper, we also asked each student to draw a picture representing another lab of their choosing, and most chose their home field of biology. We plan to report on this comparison in a future paper.

Our observation that subnetworks differed consistently in their depiction of ``Group Members,'' ``Self,'' ``Computer,'' ``Instructor,'' and ``Collaborate'' suggests attending closely to these elements in future studies. With a focus on a smaller subset of elements, one could code their depictions with greater detail, such as ideas or feelings they seem particularly connected to, or their relative size in the drawings \cite{merriman2006using,thomas1998drawing,jolley2001croatian,aronsson1996social}. For example, we could investigate the degree to which an instructor is depicted as an authority figure, as discussed in Section \ref{sec:interactions}. We could also further develop our coding scheme and perhaps add to the survey prompts to capture more detail about students' sense of their positioning within the COP. 

The approach outlined in this paper can also be paired with other means of studying student lab experiences. For example, several studies \cite{commeford2021characterizing,sundstrom2022examining,brewe2012investigating} have used social network analysis to create network diagrams of interactions between students. Similarly, Wan \textit{et al}. \cite{wan2024characterizing} characterized lab groups based on their primary discourse styles along on-task and social dimensions. Such characterizations could be compared against a drawing element network to identify how students' interactions relate to their perceptions of the lab community. The drawing survey can also be administered alongside any number of instruments from the PER literature to create subnetworks based on learning gains, attitudinal shifts, or measures of STEM identity.

Instructionally, this drawing survey and network analysis can provide feedback to course designers and instructors. Irving, McPadden, and Caballero \cite{irving2020communities} make the case that the COP framework can be a powerful curriculum design theory for introductory physics labs. The approach presented here can be used to evaluate the degree to which such a course design is internalized by students. For example, after noticing the inconsistency of goals across the network reported here, in the following semester, the course instructors added more specific life science applications to some of the more abstract studio activities and began discussing activity-specific learning goals at the beginning of each studio session. These new practices support the idea that ``learning occurs most effectively when students can articulate why what they are learning matters to them'' \cite{crouch2014introductory} in the physics community and back in their major community.

Our network analysis also identified a potential need to attend to different identity groups' experiences. We found consistent differences between identity groups' centrality for lab group membership, computers, the instructor, and collaboration. For example, female students made collaboration more central and male students made interacting with the instructor more central. This difference could be attributable to influences such as student preferences (female students preferring collaboration and male students preferring instructor feedback) or inequities (instructors unknowingly engaging more with male students). Regardless of their source, such differences provide valuable feedback regarding the course environment.

\section{Conclusions\label{sec:conclusions}}

We have demonstrated the use of a drawing-based survey and network analysis to study the elements of students' perceptions of a studio-format IPLS sequence. We used each element's frequency (how many students depicted it) and betweenness (its centrality to the network determined by weighted connections) to quantify its positioning within these perceptions. We used a bootstrapping procedure to evaluate effect sizes for betweenness values of different identity groups based on gender, college generation, and a subset of racial backgrounds. All identity groups depicted a statistically similar number of elements in their drawings and our bootstrapping procedure produced convergence and significant $p$ values, giving us confidence in making comparisons based on these effect sizes. All identity groups tended to depict practice- and member-related elements most centrally, with little consensus among their sense of the goals for the IPLS studio. 

We found that female students tended to centralize depictions of themselves, their lab group members, collaboration, and being confused or stuck, while male students tended to centralize depictions of the instructor and doing math. Continuing-generation students tended to centralize depictions of writing, learning, computers, mathematics, and collaboration, while first-generation students tended to centralize depictions of themselves and the instructor. Some of these differences persisted when we further subdivided identity groups by gender and college generation. When comparing drawings from Asian, Hispanic, and White students, we found differences in how they centralized depictions of themselves, their lab partners, collaborating and communicating, being confused, the instructor, lab equipment, computers, learning, mathematical tasks, positive feelings, questions, and writing. We demonstrated how these differences between identity groups can be compared with other studies, finding some alignment with other gender-based observations. 

We find these differences to offer important insights into student groups' experiences in the introductory lab with implications for instruction. Groups who tend to centralize depictions of themselves and their lab partners could be experiencing different senses of identity within the course or within STEM overall. Groups who tend to centralize depictions of the instructor might identify with the instructor more than others, or they might place value on the instructor's role as an authority figure. Differences in the centrality of writing, computers, and lab equipment could point to inequities found in observational studies, with implications for how students see themselves as members of the STEM community. 

In conclusion, we find that network analysis can help us visually and quantitatively represent qualitative data from students' drawn expressions of their perceptions of an introductory lab. This approach offers a holistic view of students' perceptions with important implications for research and instruction. 

\acknowledgements{This work was supported by the University of North Florida Office of Undergraduate Research. Publication was supported by a University of North Florida Faculty Publishing Grant. 
We are grateful to the UNF Astrophysics group for time on their JupyterHub server, and to J. Caleb Speirs in the PER@UNF group for his feedback on the network analysis process.}

\appendix* \section{\label{app:centrality}Other Centrality Measures}

\begin{figure*}[ht]
    \centering
    \includegraphics[width=1\textwidth]{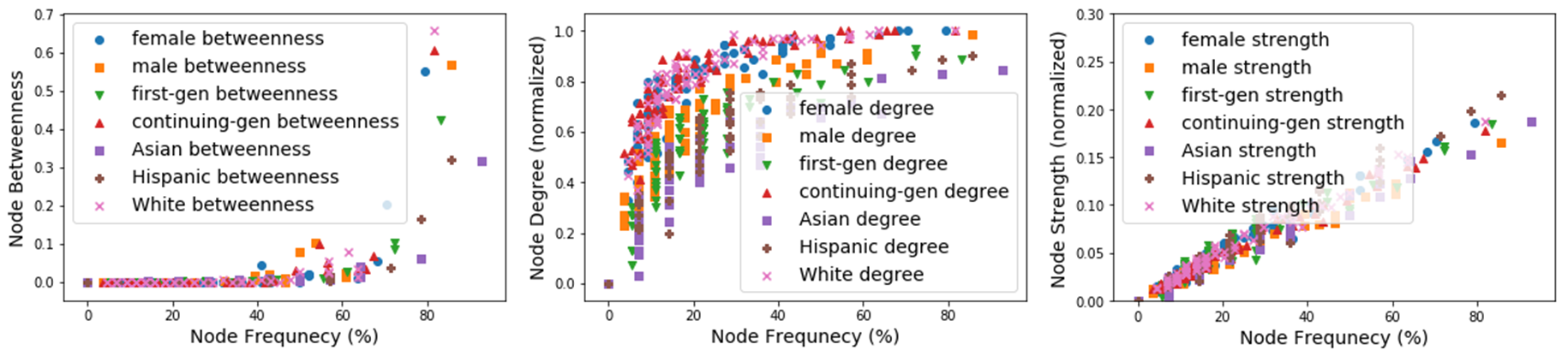}
    \caption{Centrality measures (betweeenness, normalized degree, and normalized strength) versus frequency for each node in the subnetworks considered in this paper. Betweenness shows only medium correlation with node frequency (0.51 to 0.59) while degree and strength show strong correlation (0.79 to 0.99). These correlation coefficients informed our decision to focus exclusively on betweenness in our network analysis.}
    \label{fig:centrality-measures}
\end{figure*}

Here we discuss additional centrality measures we considered in our network analysis and why we chose to omit them from discussion. These centrality measures are normalized degree, normalized strength, and closeness.

The \textit{normalized degree} $d_{i}^{\alpha}$ of node $i$ in network $\alpha$ is defined as the number of edges that connect $i$ to another node, divided by the number of other nodes in the network $N_{n}^{\alpha}-1$. Defining $\mathbbm{1}(x)$ as $1$ when $x>0$ and $0$ otherwise, 

\begin{equation}
    d_{i}^{\alpha} = \frac{\sum_{j \neq i} \mathbbm{1}\left(a_{ij}^{\alpha}\right)}{N_{n}^{\alpha}-1}.
\end{equation}

Degree is a measure of how many connections a node has within the network.

Similarly, a node's \textit{normalized strength} $s_{i}^{\alpha}$ is defined as the total weight of all the edges that connect $i$ to another node in network $\alpha$:

\begin{equation}
    s_{i}^{\alpha} = \frac{\sum_{j \neq i} a_{ij}^{\alpha}}{(N_{n}^{\alpha}-1)N_{d}^{\alpha}}.
\end{equation}

Strength is a measure of how robust a node's connections are to the rest of the network.

Finally, a node's \textit{closeness} $c_{i}^{\alpha}$ is a measure of the node's proximity to other nodes in network $\alpha$, calculated as 

\begin{equation}
    c_{i}^{\alpha} = \frac{N_{d}^{\alpha}-1}{\sum_{j} l_{ij}^{\alpha}},
\end{equation}
where $l_{ij}^{\alpha}$ is the shortest distance between nodes $i$ and $j$ in network $\alpha$, using the inverse edge weight as the distance between two connected nodes.

In this analysis, we focus on betweenness for two reasons: First, when reviewing differences in closeness values between subnetworks, we obtained suspiciously large effect sizes (Cohen's $d$ values in the range of 40 to 60). We suspect that these overly large effect sizes are an artifact of our network's sparse structure, as the full network of all drawing elements includes only $26\%$ of the total edges it could have. Additionally, some lower-frequency nodes appear in one subnetwork but not in another. A slight difference in connectivity between any two nodes could mean drastically different distances between any two other nodes in a subnetwork. Therefore, we do not report any results related to closeness.

Second, in reviewing each node's normalized degree and normalized strength within the full network of drawings and for each subnetwork that we consider in this paper, we found that both these centrality measures essentially correlated with node frequency. Figure \ref{fig:centrality-measures} shows a plot of each node's betweenness, normalized degree, and normalized strength versus node frequency for each subnetwork. In each subnetwork, normalized degree and normalized strength have a strong correlation with node frequency (correlation coefficients between 0.79 and 0.99), while betweenness has at most a medium correlation with node frequency (correlation coefficients between 0.51 and 0.59). Since we want to explore each node's importance within the network beyond the number of drawings it occurs in, we choose to focus on betweenness as the quantity most distinct from frequency.

We think this choice is reasonable, since betweenness quantifies how directly a node is connected with the rest of the network, and how often it serves as a mediator between two other nodes. This conception of centrality aligns with our earlier discussion about elements occurring together to communicate an overall idea. Normalized degree and normalized strength do not incorporate node connections beyond a given node's neighbors. Additionally, Evans and Chen \cite{evans2022linking} argue that using both betweenness and closeness may be redundant. %Even if betweenness is still a proxy for frequency in our network, the bootstrapping procedure gives us a measure of uncertainty on betweenness that is inaccessible for the node frequencies. 
Therefore, betweenness seems sufficient to help us evaluate which nodes hold important differences between two subnetworks.

\bibliography{apssamp}% Produces the bibliography via BibTeX.

\end{document}